\documentclass[
nofootinbib,
reprint,
superscriptaddress,
pra
]{revtex4-2}
\usepackage{physics}
\usepackage{bm}
\usepackage[T1]{fontenc}
\usepackage{amsmath,amsfonts,bbm, graphicx, color}
\usepackage{amssymb}
\usepackage{appendix}
\usepackage[colorlinks]{hyperref}
\usepackage[figure,table]{hypcap}
\usepackage{mathtools}

\usepackage{relsize}
\usepackage{xcolor}
\usepackage{scalerel}
\usepackage{tikz}
\usetikzlibrary{svg.path}
\usepackage{siunitx}

\definecolor{orcidlogocol}{HTML}{A6CE39}
\tikzset{
  orcidlogo/.pic={
    \fill[orcidlogocol] svg{M256,128c0,70.7-57.3,128-128,128C57.3,256,0,198.7,0,128C0,57.3,57.3,0,128,0C198.7,0,256,57.3,256,128z};
    \fill[white] svg{M86.3,186.2H70.9V79.1h15.4v48.4V186.2z}
                 svg{M108.9,79.1h41.6c39.6,0,57,28.3,57,53.6c0,27.5-21.5,53.6-56.8,53.6h-41.8V79.1z M124.3,172.4h24.5c34.9,0,42.9-26.5,42.9-39.7c0-21.5-13.7-39.7-43.7-39.7h-23.7V172.4z}
                 svg{M88.7,56.8c0,5.5-4.5,10.1-10.1,10.1c-5.6,0-10.1-4.6-10.1-10.1c0-5.6,4.5-10.1,10.1-10.1C84.2,46.7,88.7,51.3,88.7,56.8z};
  }
}

\newcommand\orcid[1]{\href{https://orcid.org/#1}{\mbox{\scalerel*{
\begin{tikzpicture}[yscale=-1,transform shape]
\pic{orcidlogo};
\end{tikzpicture}
}{|}}}}

\hypersetup{
	bookmarksnumbered,
	pdfstartview={FitH},
	citecolor={blue},
	linkcolor={blue},
	urlcolor={blue},
	pdfpagemode={UseOutlines}}
\definecolor{darkgreen}{RGB}{20,100,20}
\definecolor{darkblue}{RGB}{0,0,130}
\definecolor{darkred}{rgb}{.8,0,0}

\usepackage{soul}

\newcommand{\nn}{\nonumber}		

\usepackage{colonequals} 				

\begin{document}

\title{Quantum time dilation in atomic spectra} 

\author{Piotr T. Grochowski\orcid{0000-0002-9654-4824}}
\email{piotr@cft.edu.pl}
\affiliation{Center for Theoretical Physics, Polish Academy of Sciences, Aleja Lotnik\'ow 32/46, 02-668 Warsaw, Poland}

\author{Alexander R. H. Smith\orcid{0000-0002-4618-4832}}
\email[]{arhsmith@anselm.edu }
\affiliation{Department of Physics, Saint Anselm College, Manchester, New Hampshire 03102, USA} \affiliation{Department of Physics and Astronomy, Dartmouth College, Hanover, New Hampshire 03755, USA}

\author{Andrzej Dragan\orcid{0000-0002-5254-710X}}
\email{dragan@fuw.edu.pl}
\affiliation{Institute of Theoretical Physics, University of Warsaw, Pasteura 5, 02-093 Warsaw, Poland}
\affiliation{Centre for Quantum Technologies, National University of Singapore, 3 Science Drive 2, 117543 Singapore, Singapore}

\author{Kacper D\k{e}bski\orcid{0000-0002-8865-9066}}
\email{kdebski@fuw.edu.pl \\}
\affiliation{Institute of Theoretical Physics, University of Warsaw, Pasteura 5, 02-093 Warsaw, Poland}

\date{\today}

\begin{abstract}
Quantum time dilation occurs when a clock moves in a superposition of  relativistic momentum wave packets. The lifetime of an excited hydrogen-like atom can be used as a clock, which we use to demonstrate how quantum time dilation manifests in a spontaneous emission process. The resulting emission rate differs when compared to the emission rate of an atom prepared in a mixture of momentum wave packets at order $v^2/c^2$. This effect is accompanied by a quantum correction to the Doppler shift due to the coherence between momentum wave packets. This quantum Doppler shift affects the spectral line shape at order $v/c$. However, its effect on the decay rate is suppressed when compared to the effect of quantum time dilation. We argue that spectroscopic experiments offer a technologically feasible platform to explore the effects of quantum time dilation.
\end{abstract} 
\maketitle

\section{Introduction}

The quintessential feature of quantum mechanics is the superposition principle. When combined with relativistic effects, this principle gives rise to a number of exciting and novel phenomena~\cite{violaTestingEquivalencePrinciple1997,peresQuantumInformationRelativity2004,chiribellaQuantumComputationsDefinite2013, lorekIdealClocksConvenient2015a,kovachyQuantumSuperpositionHalfmetre2015,pikovskiUniversalDecoherenceDue2015, ruizEntanglementQuantumClocks2017, lockRelativisticQuantumClocks2017a, bassiGravitationalDecoherence2017,boseSpinEntanglementWitness2017,marlettoGravitationallyInducedEntanglement2017, pieriniCanChargedDecaying2018, zychQuantumFormulationEinstein2018, hoehnHowSwitchRelational2018, anastopoulosEquivalencePrincipleQuantum2018a, zychGravitationalMassComposite2019a,smithQuantizingTimeInteracting2019, khandelwalGeneralRelativisticTime2019, lockQuantumClassicalEffects2019a, hohnSwitchingInternalTimes2019a, zychBellTheoremTemporal2019a, Dragan2020,giacominiQuantumMechanicsCovariance2019, castro-ruizQuantumClocksTemporal2020,hoehnTrinityRelationalQuantum2019, hendersonQuantumTemporalSuperposition2020, barbadoUnruhEffectDetectors2020, fooUnruhdeWittDetectorsQuantum2020}. In particular, it is natural to ask whether there is a quantum contribution to the time dilation observed by a clock moving in a superposition of relativistic speeds. This question has been examined in several contexts: a modified twin-paradox in which one twin is placed in a superposition of motions~\cite{vedralSchrodingerCatMeets2008}; an analogue twin-paradox scenario in superconducting circuits~\cite{lindkvistTwinParadoxMacroscopic2014}; interferometry experiments in which a clock experiences a superposition of proper times~\cite{zychQuantumInterferometricVisibility2011a,bushevSingleElectronRelativistic2016,lorianiInterferenceClocksQuantum2019a, rouraGravitationalRedshiftQuantumClock2020a}; and sequential boosts of quantum clocks mimicking a twin-like scenario have been shown to lead to nonclassical effects in ion trap atomic clocks~\cite{paigeClassicalNonclassicalTime2020}. 

Recently, a probabilistic formulation of relativistic time dilation observed by quantum clocks was developed~\cite{Smith2020}. It was shown that a clock moving in a localized momentum wave packet observes on average classical time dilation in accordance with special relativity. However, the time dilation observed by a clock moving in a coherent superposition of two momentum wave packets experiences quantum correction compared to a classical clock moving in a probabilistic mixture of the same two wave packets. This quantum time dilation effect was established for an idealized model of a clock, and it thus remains open as to whether quantum time dilation is universal, analogous to the way in which classical time dilation affects all clocks in the same way.

The purpose of the present work is to provide evidence in support of the conjecture that quantum time dilation is universal.
We consider the lifetime of an excited hydrogen-like atom as a clock~\cite{peresMeasurementTimeQuantum1980} and demonstrate that when such an atom moves in a coherent superposition of momenta its lifetime experiences the same quantum time dilation as the clocks considered in~\cite{Smith2020}.
This yields a spectroscopic signature of a clock experiencing a superposition of proper times, alternative to past interferometry proposals that aim to observe a decrease in interference visibility~\cite{zychQuantumInterferometricVisibility2011a,bushevSingleElectronRelativistic2016,lorianiInterferenceClocksQuantum2019a, rouraGravitationalRedshiftQuantumClock2020a}.

Spectroscopic signatures of classical time dilation have been observed for atoms moving as speeds as low as \SI{10}{m/s}~\cite{chouOpticalClocksRelativity2010}. Nonclassical effects in emission spectroscopy due to the coherent spreading of the atomic center-of-mass wave function were first studied in the early 1990s \cite{Bialynicka1991,Rzazewski1992,Steuernagel1995,Fedorov2005}, and the effect of center-of-mass superposition was recently investigated in a scalar field model~\cite{stritzelbergerCoherentDelocalizationLightmatter2020}.
In the present work, we show that the exact quantum time dilation effect described in~\cite{Smith2020} is observed in the spontaneous decay rate of an atom moving in a coherent superposition of relativistic momenta. This observation motivates a new class of spectroscopic measurements that are sensitive to relativistic effects due to quantum coherence.

In addition, a novel correction to the classical Doppler shift is shown to modify the shape of the atomic emission spectrum. This correction is present for light emitted in the direction of the atom's motion. On the other hand, if the spectrum is measured through photons perpendicular to motion, effects that are first-order in momentum  vanish and give way to second-order, relativistic corrections. This is clearly seen from the angular distributions of radiation coming from moving atoms which we present. These distributions show the directions of emission that are the most affected by motion and suggest the optimal way to measure the quantum time dilation.

Furthermore, we analyze potential experimental scenarios in which both the quantum Doppler and the quantum time dilation effects can be measured.
As of now, spectroscopic experiments have been able to observe classical time dilation in atoms moving as slowly as $10$ m/s~\cite{chouOpticalClocksRelativity2010}.
Basing on that and the subsequent advances, we argue that state-of-the-art techniques involving atomic ion clocks and large momentum transfer setups can reach necessary parameter regimes.

\section{Quantum time dilation}
\label{Quantum time dilation}
As a model of a quantum clock, consider a relativistic particle with an internal degree of freedom, described by the Hamiltonian
\begin{equation}
    \hat{H} = \sqrt{  \hat{\boldsymbol{p}}^2 c^2 + \hat{M}^2 c^4 }, \label{relativisticInternalClockHamiltonian}
\end{equation}
where $\hat{\boldsymbol{p}}$ is the particle's momentum and $\hat{M} \equiv  m + \hat{H}_{\rm clock}/c^2$ is the so called mass operator, which is a combination of the particle's rest mass $m$ and the dynamical mass $\hat{H}_{\rm clock}/c^2$ stemming from the energy of the internal degree of freedom governed by the Hamiltonian $\hat{H}_{\rm clock}$. This internal degree of freedom can serve as a clock that tracks the particles proper time as measured by a time observable $T_{\rm clock}$ that transforms covariantly with respect to the group generated by $\hat{H}_{\rm clock}$;\footnote{In more detail, the time observable $T_{\rm clock}$ is defined as a positive operator valued measure with effect operators $\hat{E}(t)$ that satisfy the covariance condition $\hat{E}(t+t') = e^{-it'\hat{H}_{\rm clock}} \hat{E}(t) e^{it'\hat{H}_{\rm clock}}$~\cite{holevoProbabilisticStatisticalAspects1982,braunsteinGeneralizedUncertaintyRelations1996, buschTimeObservablesQuantum1994}. We make the further assumption that this time observable is sharp and can thus be associated with a self-adjoint operator $\hat{T}_{\rm clock}$.} such covariant observables are common in parameter estimation tasks and in this case gives the best estimate of the proper time experienced by the particle.

Consider the particle to be prepared in a superposition of momentum wave packets, which up to normalization is taken to be
\begin{equation} \label{superpos}
    \ket{\psi} \sim \cos \theta \ket{\varphi_{\bar{\boldsymbol{p}}_1}} + e^{i \phi}\sin \theta \ket{\varphi_{\bar{\boldsymbol{p}}_2}},
\end{equation}
where $\theta \in [0,\frac{\pi}{2})$, $\phi \in [0,\pi)$, and $\braket{\bm{p}}{ \varphi_{\bar{\bm{p}}_i}}= e^{-(\boldsymbol{p}-\bar{\boldsymbol{p}}_i)^2/2 \Delta^2}/\pi^{1/4}\sqrt{\Delta}$ with $\Delta$ being the spread of the wave packet in momentum space. Let the clock be characterized by the Hilbert space $L^2(\mathbb{R})$, with the Hamiltonian equal to the momentum operator, $\hat{H}_{\rm clock} = c \hat{P}_{\rm clock}$, and the covariant time observable satisfies $[\hat{T}_{\rm clock}, \hat{H}_{\rm clock}] = i \hslash$. The average time read by the clock $\langle \hat{T}_{\rm clock}\rangle$ when the clock of an observer relative to which the Hamiltonian in Eq.~\eqref{relativisticInternalClockHamiltonian} generates an evolution reads the time $t$  can be shown to be equal to~\cite{Smith2020}
\begin{equation}
    \langle \hat{T}_{\rm clock}\rangle = \left( \gamma_{\rm C}^{-1} + \gamma_{\rm Q}^{-1}\right) t,
    \label{quantumTimeDilation}
\end{equation}
where to leading relativistic order
\begin{align}
    \gamma_{\rm C}^{-1} \equiv 1- \frac{ \bar{\boldsymbol{p}}_1^2 \cos^2 \theta +  \bar{\boldsymbol{p}}_2^2 \sin^2 \theta - \Delta^2/2}{2m^2 c^2 },
    \label{gammaC}
\end{align}
is associated with the classical time dilation of a clock moving in a statistical mixture of momenta $\bar{\boldsymbol{p}}_1$ and $\bar{\boldsymbol{p}}_1$ with probabilities $\cos^2 \theta$ and $\sin^2 \theta$, and\\
\begin{equation}
    \gamma_{\rm Q}^{-1} \equiv \frac{ \cos \phi \sin 2 \theta \! \left[ \left(\bar{ \boldsymbol p}_2-\bar{ \boldsymbol p}_1\right)^2-2 \!\left(\bar{\boldsymbol p}_2^2-\bar{\boldsymbol p}_1^2\right) \! \cos 2 \theta \right] }{8 m^2 c^2 \left[\cos \phi \sin 2 \theta +e^{\frac{( \bar{\boldsymbol p}_2 - \bar{\boldsymbol p}_1 )^2}{4 \Delta^2}}\right] },
    \label{gammaQ}
\end{equation}
 quantifies corrections to classical time dilation resulting from coherence across the momentum wave packets carrying the internal clock. Equation~\eqref{quantumTimeDilation} can be thought of as the generalization of the classical time dilation formula that takes into account the possibility of the clock moving in nonclassical states of motion and nonzero $\gamma_{\rm Q}^{-1}$ leads to quantum time dilation effects.

The above considerations were based on an ideal clock model in which the proper time of the clock was associated with an operator that was canonically conjugate to the clock Hamiltonian. It is thus not clear whether quantum time dilation is universal, affecting all clocks in the same way analogous to its classical counter part, and the answer must ultimately come from experiment.

In what follows we present evidence that supports the conjecture that quantum time dilation between clocks moving in superpositions of inertial trajectories is universal. We consider an entirely different clock model based on the lifetime of an excited atom and demonstrate that such a clock observes quantum time dilation in accordance with Eq.~\eqref{gammaQ}. This brings quantum time dilation effects closer to experiment by demonstrating that they manifest in a realistic clock model based on spontaneous emission, the mechanism by which atomic clocks operate.

\section{Spectroscopy of moving atoms}

Consider a two-level atom of mass $m$ and suppose that its ground state $\ket{g}$ and excited state $\ket{e}$ are separated by an energy difference $\hslash\Omega$ in the atom's rest frame. The dynamics of the atom and electromagnetic fields, $\hat{\boldsymbol{E}}$ and $\hat{\boldsymbol{B}}$, are described by the Hamiltonian
\begin{equation}\label{ham}
\hat{H} = \hat{H}_\text{atom} +\hat{H}_\text{field}+\hat{H}_\text{\text{af}}, 
\end{equation}
where the free Hamiltonian of the atom to leading relativistic order in the atom's center-of-mass momentum $\hat{\bm{p}}/mc$~(e.g.\ \cite{zychQuantumSystemsGravitational2017}) is
\begin{equation}
\hat{H}_\text{atom}= \frac{\hat{\bm{p}} ^2}{2 m} -\frac{1}{8} \frac{\hat{\bm {p}}^4}{ m^3 c^2} + \hslash \Omega \left( 1 - \frac{1}{2} \frac{\hat{\bm {p}}^2}{ m^2 c^2} \right) \ket{e}\!\bra{e},
\end{equation}
and the electromagnetic field Hamiltonian is  $\hat{H}_\text{field} = \sum_{\bm{k}, \xi} \hslash \omega_k \hat{a}^{\dagger}_{\bm{k}, \xi} \hat{a}_{\bm{k}, \xi}$, which is a mode sum over the wave vector $\bm{k}$ and polarization index $\xi$ with the corresponding eigenfrequencies $\omega_k=kc$ and annihilation operators $\hat{a}_{\bm{k}, \xi}$. The atom is coupled to the electromagnetic field through the interaction Hamiltonian \cite{Wilkens1993,Wilkens1994,barnettVacuumFrictionParadox2018,Sonnleitner2017,Sonnleitner2018}
\begin{equation}\label{inter1}
\hat{H}_\text{\text{af}} = -\hat{\bm{d}} \cdot \hat{\bm{E}}^{\perp}-\frac{1}{2 m} \left[ \hat{\bm{p}} \cdot 
\left( \hat{\bm{B}} \times \hat{\bm{d}} \right) + \left( \hat{\bm{B}} \times \hat{\bm{d}} \right) \cdot \hat{\bm{p}} \right],
\end{equation}
where the first term is the usual dipole interaction, $\hat{\bm{d}} = \bm{d}\left( \ket{g}\!\bra{e}+\ket{e}\!\bra{g}\right)$ the dipole operator in the lab frame, and the second term is the so-called R\"ontgen term that accounts for the Lorentz-transformed electromagnetic field felt by the moving atom.\footnote{This term has been known since the 19th century~\cite{Rontgen1888}, however it was omitted in the early works on light-matter interactions. One of its first incorporations dates back to works of Babiker in the 1980s~\cite{Babiker1984}, yet it wasn't until the early 2000s that this term was rigorously shown to be necessary to retrieve agreement with special relativity~\cite{Wilkens1994,Boussiakou2002,cresserRateSpontaneousDecay2003}.
Nowadays, it is routinely utilized in studies of atom-field interactions involving moving bodies~\mbox{\cite{Matloob2005,Sonnleitner2017,sonnleitnerMassenergyAnomalousFriction2018,schwartzPostNewtonianHamiltonianDescription2019b,shafieiyanSpontaneousEmissionMoving2018}.} The R\"ontgen term is a consequence of the vectorial nature of the electromagnetic field, and thus would not appear in an analogous scalar field model~\cite{stritzelbergerCoherentDelocalizationLightmatter2020,lopp2020quantum}.} It is important to note that all the operators entering the Hamiltonian~\eqref{ham} 
 are expressed in the laboratory frame.\footnote{For example, the dipole moment is connected to its rest value $\boldsymbol{d}'$ by a Lorentz transformation $\boldsymbol{d} = \boldsymbol{d'}-\frac{{\boldsymbol d'}\cdot{\boldsymbol v}}{v^2}{\boldsymbol v} +
\frac{{\boldsymbol d'}\cdot{\boldsymbol v}}{v^2}{\boldsymbol v} / \sqrt{1-v^2/c^2}$, where $\boldsymbol{v}$ is the velocity of the moving atom.} The electromagnetic fields appearing in \eqref{inter1} are given by
\begin{align}
\hat{\bm{E}}^{\perp} (\bm{r}) &= - i \sum_{\bm{k}, \xi} \sqrt{\frac{\hslash \omega_k}{2 \epsilon_0 V}} \bm{\epsilon}_{\bm{k}, \xi} \hat{a}_{\bm{k}, \xi} e^{i \bm{k} \cdot \hat{\bm{r}}} + \text{H.c.}, \\
\hat{\bm{B}} (\bm{r}) &= i \sum_{\bm{k}, \xi} \sqrt{\frac{\hslash }{2 \epsilon_0 V \omega_k}} \left( \bm{k} \times \bm{\epsilon}_{\bm{k}, \xi}\right) \hat{a}_{\bm{k}, \xi} e^{i \bm{k} \cdot \hat{\bm{r}}} + \text{H.c.}, 
\end{align}
where $\epsilon_0$ is the vacuum permittivity and $V$ the quantization volume, while $\bm{\epsilon}_{\bm{k}, \xi}$ is the polarization vector perpendicular to the wave vector $\bm{k}$. By invoking the rotating wave approximation~\cite{Wilkens1994, Sonnleitner2017}, the interaction Hamiltonian~\eqref{inter1} assumes the form
\begin{equation}\label{inter2}
\hat{H}_\text{\text{af}} = -i\hslash \sum_{\bm{k}, \xi} \sqrt{\frac{\hslash \omega_k}{2 \epsilon_0 V}} \hat{g}_{\bm{k}, \xi} e^{i \bm{k} \cdot \hat{\bm{r}}} \ket{e}\!\bra{g}\hat{a}_{\bm{k}, \xi} +\text{H.c.}, 
\end{equation}
where the coupling `constant' depends on the atom's momentum 
\begin{equation}\label{coup}
 \hat{g}_{\bm{k}, \xi} = \bm{\epsilon}_{\bm{k}, \xi} \cdot \bm{d} +\frac{1}{m \omega_k} \left( \hat{\bm{p}} - \frac{\hslash \bm{k}}{2} \right)\cdot \left[ \left( \bm{k} \times \bm{\epsilon}_{\bm{k}, \xi} \right) \times \bm{d} \right], 
\end{equation}
and is an operator itself with eigenvalues $g_{\bm{k}, \xi}$.
Note that $\hat{H}_\text{\text{af}}$ has an explicit dependence on the center-of-mass position operator $\hat{\bm{r}}$, which is treated as a quantum degree of freedom. This Hamiltonian couples the internal energy levels of the atom to the center-of-mass degree of freedom, and as a consequence causes a recoil of the decaying atom.
The second term of Eq.~\eqref{coup} is a direct result of the R\"ontgen term in Eq.~\eqref{inter1}.

Let us take a moment to consider the energy scales characterizing the situation just described, namely the energy of the atom's internal degree of freedom $\hslash \Omega$, rest energy $m c^2$, and kinetic energy  $\langle \hat{\bm{p}}^2 /2m\rangle$. In what follows we will consider regimes in which the internal energy of the atom is much smaller than both its rest energy and kinetic energy, which ensures that a first order expansion in both $\hslash \Omega / m c^2$ and $\hslash \Omega / \langle \hat{\bm{p}}^2/2m \rangle$ is valid. 

\begin{figure*}[t]
	\includegraphics[width=0.31\linewidth]{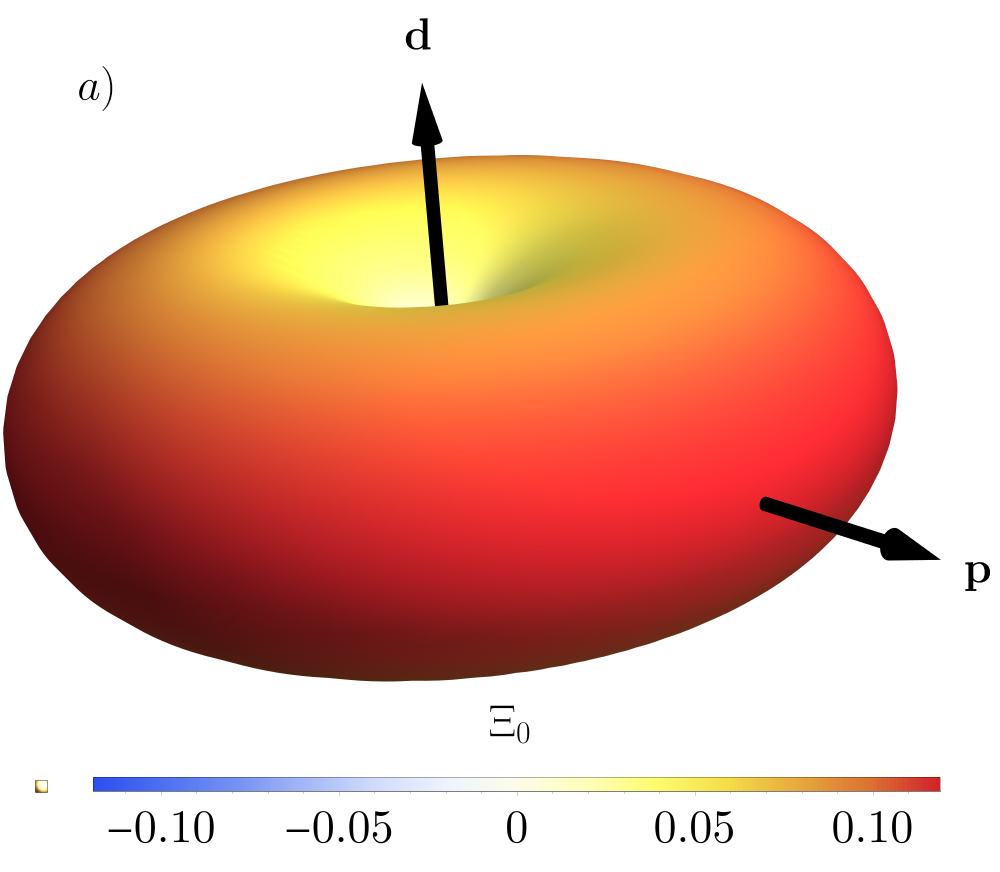}
	\includegraphics[width=0.31\linewidth]{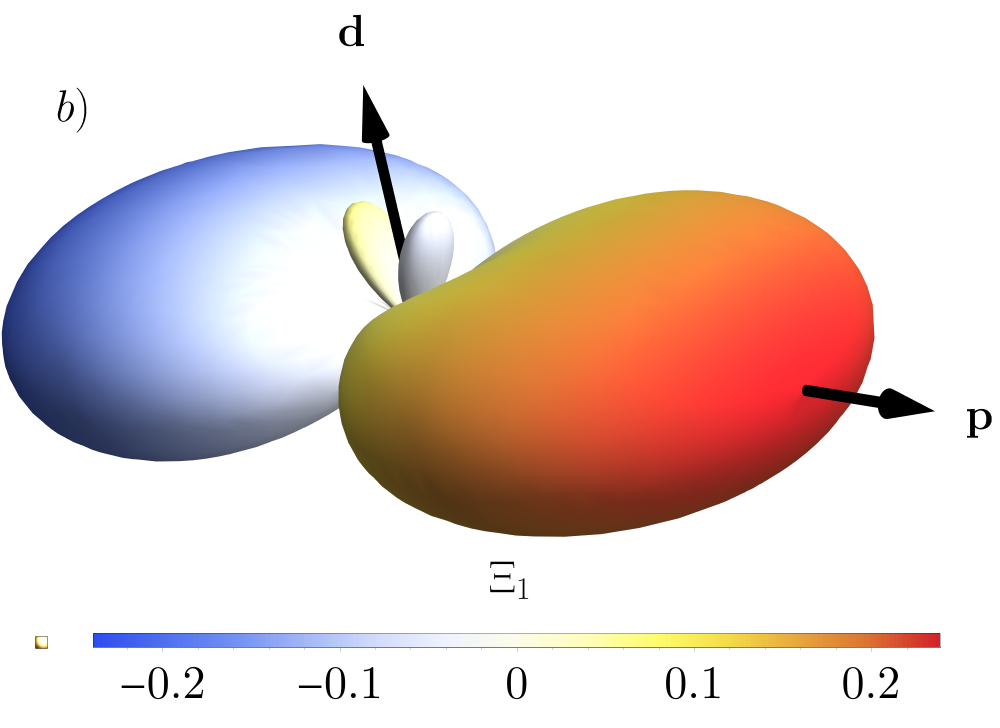}
	\includegraphics[width=0.31\linewidth]{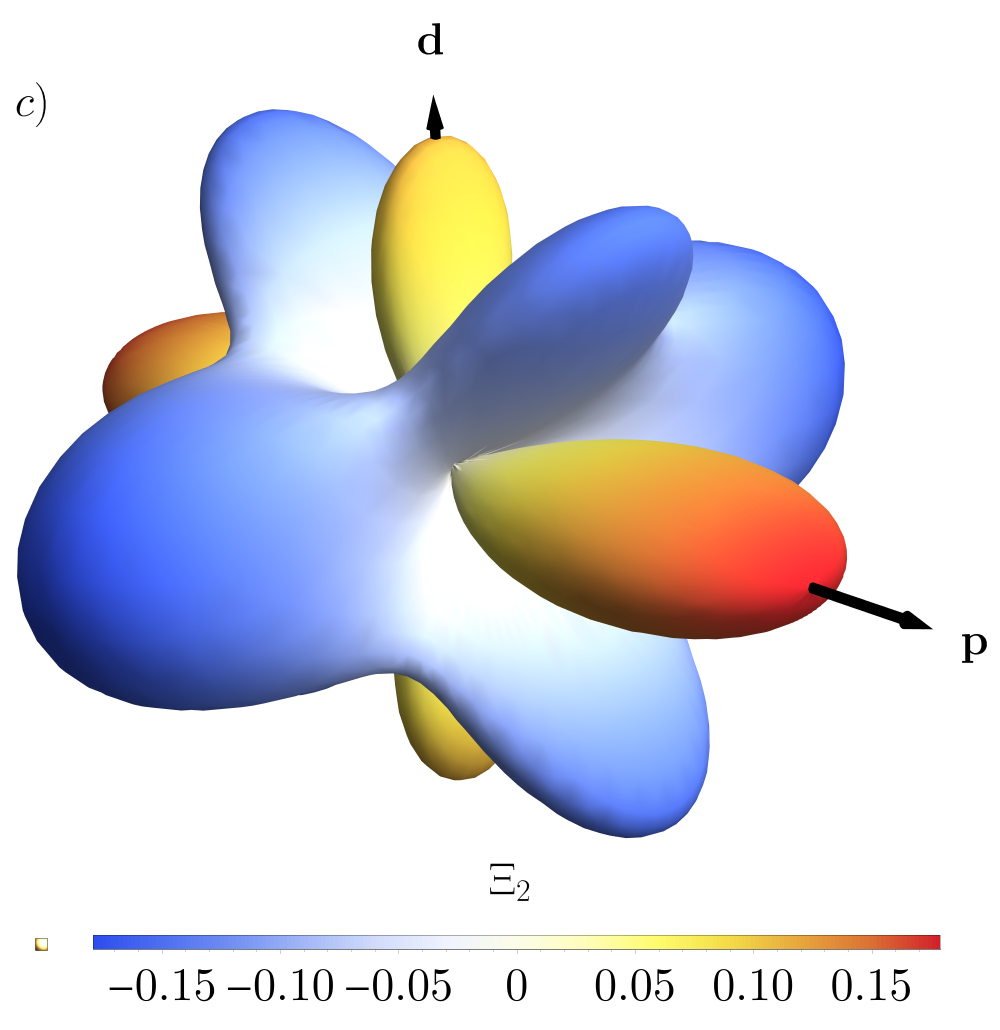}
	\caption{\label{fig0} 
a) Angular dipole distribution of emitted photons from a decaying atom at rest with respect to the center of mass momentum $\bm{p}$ and dipole momentum $\bm{d}$. Motional corrections to this angular distribution b) linear and c) quadratic in the atom's center of mass momentum. Magnitudes are represented by the distance to the origin and color (red positive, blue negative)}
\end{figure*}
Suppose an atom begins in its excited state with center-of-mass wave function $\psi(\bm{p})$ and the electromagnetic field in the vacuum, $\ket{\Psi(0)} = \int \dd \boldsymbol{p} \, \psi(\boldsymbol{p})\ket{e,\bm{p},0}$. At a later time $t$, the composite system evolves to the state
\begin{align}
\ket{\Psi (t)} &= \int \dd \bm{p} \, \alpha \left( \bm{p},t \right) \ket{e,\bm{p},0} \nonumber \\
&\quad + \sum_{\bm{k}, \xi} \int \dd \bm{p} \, \beta_{\bm{k}, \xi} \left( \bm{p},t \right) \ket{g,\bm{p}-\hslash \bm{k},1_{\bm{k}, \xi}},
\end{align}
which has been expanded in the energy eigenstates $\ket{e,\bm{p},0}$ and $\ket{g,\bm{p}-\hslash \bm{k},1_{\bm{k}, \xi}}$, associated respectively with the energies
\begin{align}
\hslash \omega_e (\bm{p}) &= \frac{\bm{p}^2}{2 m} - \frac{\bm{p}^4}{8 m^3 c^2}+ \hslash \Omega \left(1-\frac{1}{2} \frac{\bm{p}^2}{m^2 c^2}\right),  \\
\hslash \omega_g (\bm{p}, \bm{k}) &= \frac{\left( \bm{p}-\hslash \bm{k}\right) ^2}{2 m} -\frac{\left( \bm{p}-\hslash \bm{k}\right) ^4}{8 m^3 c^2}+ \hslash \omega_k. 
\end{align} 
The time-dependent coefficients in $\ket{\Psi(t)}$ can be obtained by solving the associated Schr\"{o}dinger equation via a Laplace transform as commonly utilized in Wigner-Weisskopf theory~\cite{Weisskopf1930}. Using a single pole approximation~\cite{Rzazewski1992} one finds 
\begin{align} \label{coeffs}
\alpha \left( \bm{p},t \right) &= e^{- i \omega_e (\bm{p}) t } e^{-\frac{\Gamma \left(\bm{p}\right)}{2} t} \psi \left( \bm{p} \right),  \\
\beta_{\bm{k}, \xi} \left( \bm{p},t \right) &= \sqrt{\frac{\hslash \omega_k}{2 \epsilon_0 V}} g_{\bm{k}, \xi} (\bm{p}) \psi \left( \bm{p} \right) \nn \\
&\quad \times \frac{ e^{-i \omega_e (\bm{p}) t } e^{-\frac{\Gamma \left(\bm{p}\right)}{2} t}-e^{- i \omega_g (\bm{p},\bm{k}) t } }{i \left[ \omega_e (\bm{p}) - \omega_g (\bm{p},\bm{k}) \right] + \frac{\Gamma \left(\bm{p}\right)}{2} }, 
\end{align}
where $\Gamma (\bm{p})$ is the total transition rate of the spontaneous decay of the atom moving with momentum $\bm{p}$:
\begin{equation}
 \Gamma (\bm{p})  =  \sum_{\bm{k}, \xi}  \frac{  \omega_k}{ 8 \pi^2 \hbar \epsilon_0 c^3} g_{\bm{k}, \xi}^2 (\bm{p}) \delta \left[ \omega_e (\bm{p}) - \omega_g (\bm{p},\bm{k}) \right].   
\end{equation} 
Then, the total transition rate in the long-time limit is
\begin{equation} \label{total}
 \Gamma \!=\! \lim_{t \rightarrow \infty} \frac{\dd}{\dd t} \sum_{\bm{k}, \xi} \int \dd \bm{p} \, \left| \beta_{\bm{k}, \xi} \left( \bm{p},t \right) \right|^2 
 \!=\! \int \dd \bm{p} \, \left| \psi \left( \bm{p} \right) \right|^2 \Gamma (\bm{p}),
\end{equation}
where the results of~\cite{Boussiakou2002,cresserRateSpontaneousDecay2003} are recovered when $\psi \left( \bm{p} \right)$ is a momentum eigenstate.
As detailed in the Appendix~\ref{emission_app}, the angular distribution of the emitted radiation is obtained by omitting the angular integration in Eq.~\eqref{total}, yielding
\begin{align} \label{fulang}
\frac{\Gamma(\Theta,\Phi)}{\Gamma_0} &=   \Xi_0 (\Theta,\Phi) \left( 1 - \frac{3}{2}\frac{ \hbar \Omega}{  m c^2} \right)   \nonumber \\
&\quad +  \frac{1}{m c} \Xi_1 (\Theta,\Phi) \int \dd p \ p |\psi(p)|^2  \nonumber \\
&\quad + \frac{1}{2 m^2 c^2} \Xi_2 (\Theta,\Phi)  \int \dd p \ p^2 |\psi(p)|^2 ,
\end{align}
where $\Theta$ and $\Phi$ are the azimuthal and polar angles of $\bm{k}$ vector relative to $\bm{p}$, respectively, $\Gamma_0 = \frac{\Omega^3 d'^2}{3 \pi \epsilon_0 \hslash c^3}$ is the total decay rate of a standing atom ignoring recoil effects (i.e., $\hbar \Omega \ll m c^2$), and we have assumed that the atom moves only along the $z$ axis, perpendicular to the dipole moment vector $\bm{d}$.
As such, $\psi(p)$ has to be understood as a marginal distribution of a full center-of-mass wave function, $\abs{\psi(p)}^2=\int \dd p_x \dd p_y \, \abs{\psi(\bm{p})}^2$, where momentum distributions in the $x$ and $y$ directions are well localized around the $z$ axis.
$\Xi_0 (\Theta,\Phi)$ is the standard angular distribution of dipole radiation 
\begin{align} 
\Xi_0 (\Theta,\Phi) = \frac{3}{8 \pi}  \left( 1 - \sin^2 \Theta \cos^2 \Phi\right),
\end{align}
while
\begin{align} 
\Xi_1 (\Theta,\Phi)  &= \frac{3}{4 \pi} \cos \Theta \left( 1 - 2 \sin^2 \Theta \cos^2 \Phi \right),  \\
\Xi_2 (\Theta,\Phi) &=  \frac{3}{16 \pi}  
 \left[ 6 \cos 2 \Theta + 5 \cos^2 \Phi \left( \cos 4 \Theta - \cos 2 \Theta \right)\right],  
\end{align}
are first and second order corrections in $p/mc$ to the dipole distribution appearing due to the motion of the atom~\cite{Wilkens1994}. These motional corrections to the angular distribution of radiation are universal as they manifest unless the atom is at rest and their shape is independent of the momentum wave function $\psi(p)$ (see Fig.~\ref{fig0}). 

Integration over $\Theta$ and $\Phi$ recovers the familiar formula~\cite{Wilkens1994,Boussiakou2002,cresserRateSpontaneousDecay2003,Sonnleitner2017}
\begin{align}\label{totGamma}
\Gamma 
 &= \Gamma_0 \left(1 - \frac{3 \hbar \Omega}{ 2 m c^2} - \frac{1}{2 m^2 c^2} \int \dd p \ p^2 |\psi(p)|^2  \right). 
\end{align}
If the atom were to move along a classical trajectory with momentum $\bar{p}$, corresponding $|\psi(p)|^2 = \delta(p-\bar{p})$,  the transition transition rate $\Gamma$ is related to transition rate in the atom's rest frame $\Gamma_0 \left(1 - \frac{3 \hbar \Omega}{ 2 m c^2}\right)$ via a Lorentz factor, $\Gamma = \Gamma_0 (1 - \frac{3 \hbar \Omega}{ 2 m c^2}) \sqrt{1-v^2/ c^2} \approx \Gamma_0 (1 - \frac{3 \hbar \Omega}{ 2 m c^2} - \frac{\bar{p}^2}{ 2m^2c^2} )$, which agrees with Eq.~\eqref{totGamma} and ensures consistency with special relativity.
\begin{figure*}[t]
	\includegraphics[height=0.31\linewidth]{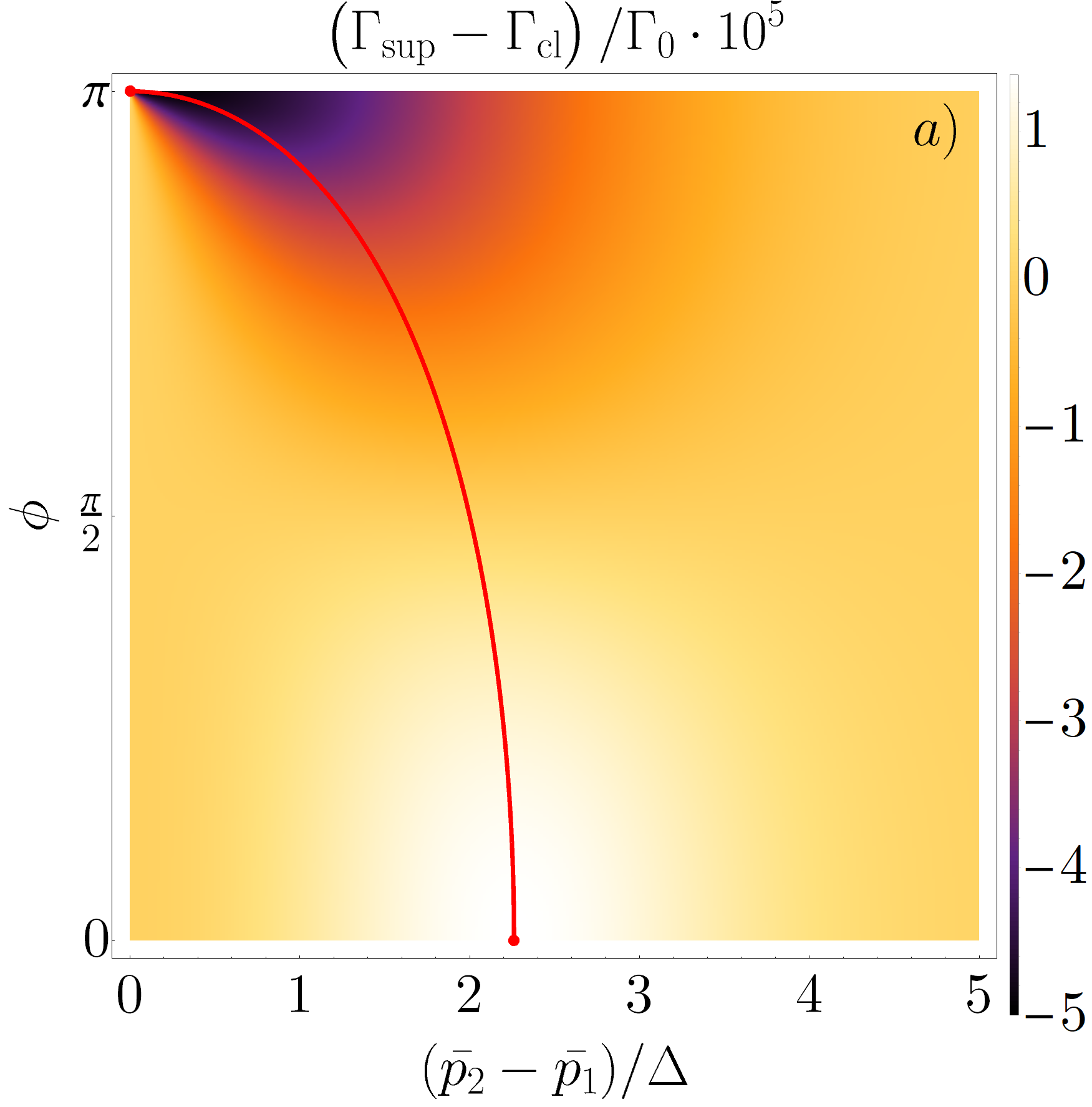} \quad 
	\includegraphics[height=0.31\linewidth]{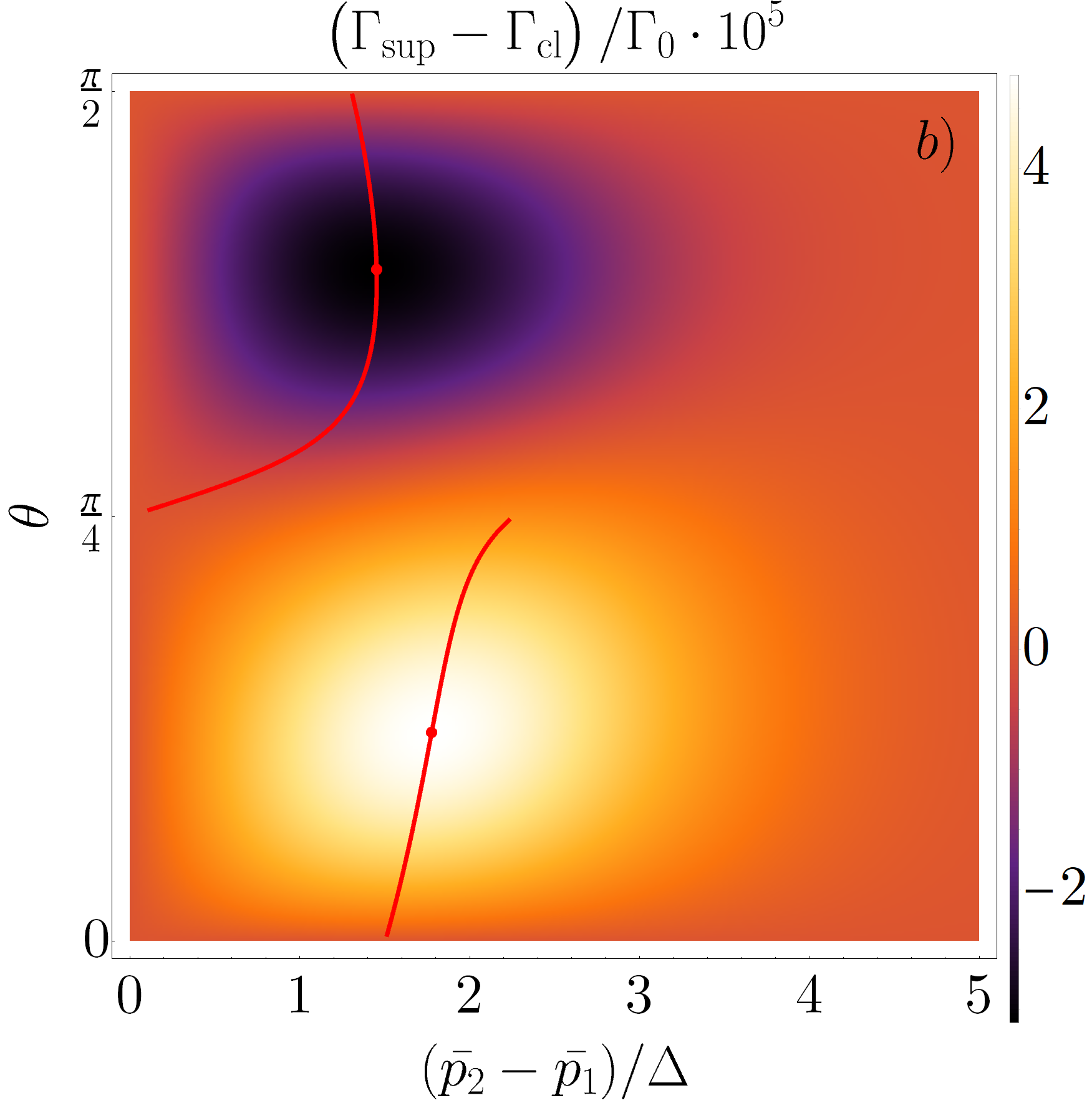} \quad 
	\includegraphics[height=0.31\linewidth]{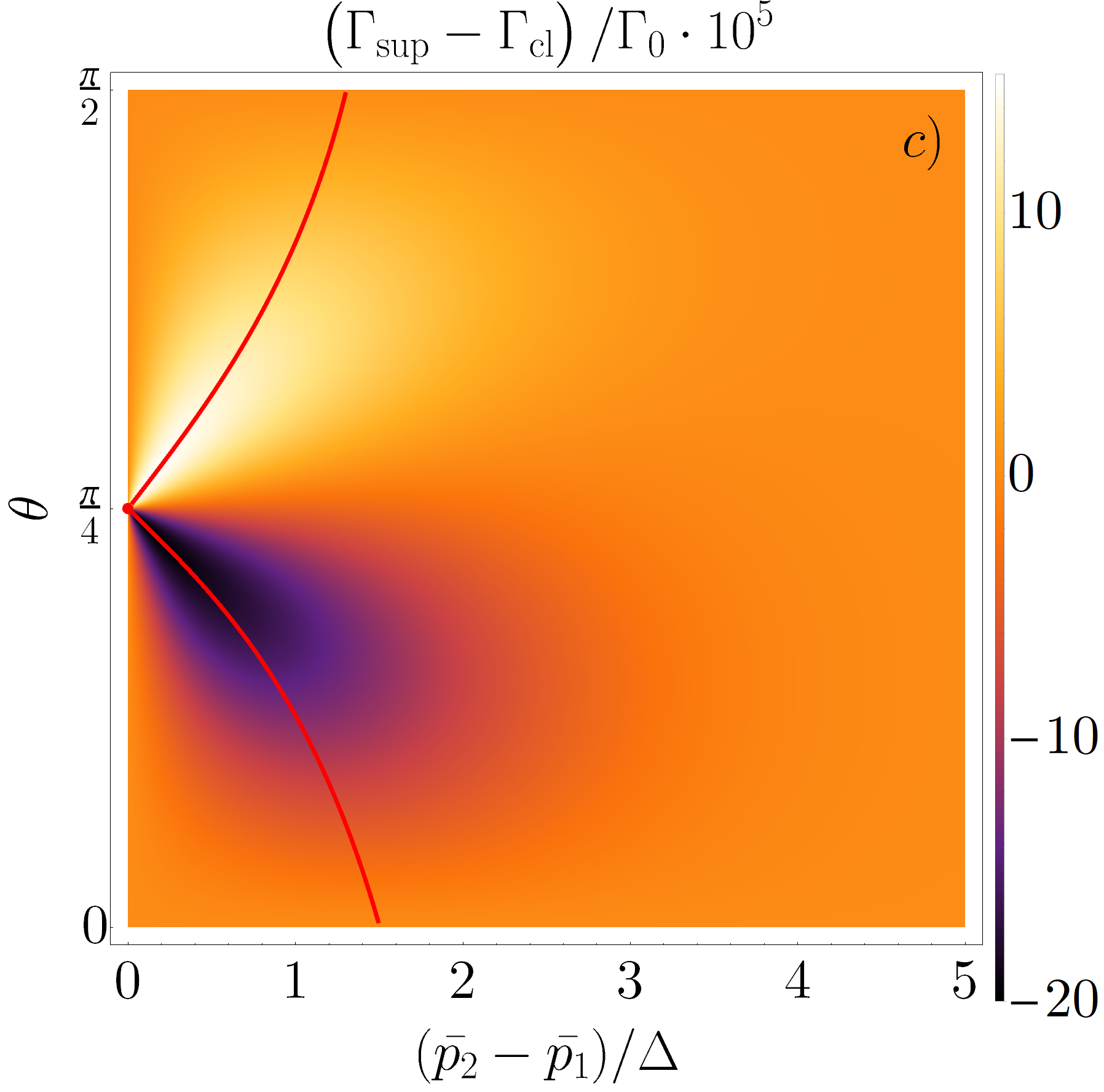}
	\caption{\label{fig1} 
	The difference in total emission rate between a superposition and a classical mixture of two momentum wave packets of an atom as a function of the wave packets' momentum difference and their relative phase and weight: a) equal weighted superposition of momentum wave packets, $\theta = \pi / 4$, b) Relative phase fixed at $\phi = 0$, c) Relative phase fixed at $\phi = \pi$.
	The red line marks the maximum value of the effect for a given relative phase or a relative weight, while the red circles signify maximum and minimum values across the whole plot.
	Nonzero value for a finite momentum difference signifies the phenomenon of quantum time dilation.
	In each of the panels, the momentum spread of each of wave packets is $\Delta = 0.01 m c$ and the sum of their average momenta is equal to $\bar{p}_1 + \bar{p}_2 = 0.05 m c$.
}
\end{figure*}

Additionally, from Eqs.~\eqref{coeffs} one can extract the shape of an emission line, which can be straightforwardly transformed to the absorption spectrum through the Einstein coefficients.
The probability that an atom emits a photon with momentum $\hslash\bm{k}$ is
\begin{align}
 \mathcal{P}\left( \bm{k} \right) &= \lim_{t \rightarrow \infty} \sum_{\xi} \int \dd \bm{p}  \, \left| \beta_{\bm{k}, \xi} \left( \bm{p},t \right) \right|^2.
\end{align}
Again, assuming that the atom has a large mass $m$, moves along the $z$ axis and dipole moment points perpendicularly to motion, we arrive at the following characteristic of a transition line for photons emitted along the direction of motion (see Appendix~\ref{emission_app}):
\begin{align}\label{tranal0}
&\mathcal{P}_{\parallel} (\omega)  =\frac{3}{8 \pi}  \times \nonumber \\
&\int \dd p \, \left| \psi \left( p \right) \right|^2 \frac{ \left( 1 + 3 \frac{ p}{m c } \right) \Gamma_0 / 2 \pi  }{\left[ \omega - \Omega \left(1+ \frac{ p}{mc}\right) \right]^2 + \frac{\Gamma_0^2}{4} \left( 1+ 2 \frac{ p}{mc} \right) },
\end{align}
and perpendicular to both the dipole moment and to the direction of motion: 
\begin{align}\label{tranal01}
&\mathcal{P}_{\perp} (\omega)  =\frac{3}{8 \pi}  \times \nonumber \\
&\int \dd p \, \left| \psi \left( p \right) \right|^2 \frac{ \left( 1 -\frac{3}{2}\frac{ p^2}{m^2 c^2 } \right) \Gamma_0 / 2 \pi  }{\left[ \omega - \Omega \left(1 - \frac{1}{2}\frac{ p^2}{m^2c^2}\right) \right]^2 + \frac{\Gamma_0^2}{4} \left( 1-  \frac{ p^2}{m^2c^2} \right) }.
\end{align}
Note that both $\mathcal{P}_{\parallel} (\omega)$ and $\mathcal{P}_{\perp} (\omega)$ have been expanded up to the leading relativistic order and are proportional to the center-of-mass momentum distribution $\abs{\psi(p)}^2$ integrated against a Lorenz distribution. When observed in the direction of motion, the transition line is Doppler shifted, as the Lorentz distribution is shifted linearly in momentum by an amount $\Omega \rightarrow \Omega (1+p/mc)$. On the other hand, light emitted/absorbed perpendicular to the motion is not affected by this Doppler shift and relativistic corrections are dominant, shifting the center of the Lorentz distribution by an amount $\Omega \rightarrow \Omega (1-p^2/2 m^2c^2)$, which is  quadratically in momentum.

Each of the quantities of interest\,---\,the angular distribution of radiation, the total decay rate, and the shape of the emission line\,---\,are routinely measured in various experiments~\cite{welz1999atomic}. As we have shown how these observables depend on the center-of-mass momentum distribution, we are now equipped to show how nonclassical center-of-mass motion in such experimental scenarios can be utilized as a direct probe of quantum time dilation.

\section{Spectroscopic signatures of quantum time dilation}

First, we will compare the transition rate $\Gamma$ between atoms in coherent superpositions and incoherent classical mixtures of localized momentum wave packets. Analogously to the quantum clock model described in Sec.~\ref{Quantum time dilation}, an atom is considered to be either in a superposition~\eqref{superpos} with a momentum distribution given by
\begin{align}
    \psi_\text{sup} (\bm{p}) =  \mathcal{N} \left(
\cos \theta \braket{\bm{p}}{ \varphi_{\bar{\bm{p}}_1}} + e^{i \phi} \sin \theta \braket{\bm{p}}{ \varphi_{\bar{\bm{p}}_2} } \right),
\end{align}
where 
\begin{equation}
\mathcal{N} = \left[ \sqrt{\pi} \Delta \left( 1+ \cos \phi \ \sin 2 \theta \ e^{ -\frac{(\bar{\bm{p}}_1-\bar{\bm{p}}_2)^2}{4\Delta^2}} \right)\right]^{-1/2},
\end{equation}
or in a classical mixture such that
\begin{equation} 
   P_{\rm cl}(\bm{p}) = \cos^2 \theta \  \left|\braket{\bm{p}}{ \varphi_{\bar{\bm{p}}_1}} \right|^2 + \sin^2 \theta \  \left|\braket{\bm{p}}{ \varphi_{\bar{\bm{p}}_2}} \right|^2
\end{equation}
of momentum wave packets. For simplicity, we consider $\bar{\bm{p}}_1$ and $\bar{\bm{p}}_2$ to be collinear. By evaluating \eqref{total} we arrive at the following relative difference of total emission rates between these two cases:
\begin{align}\label{corrections}
\frac{\Gamma_{\text{sup}}\!-\Gamma_{\text{cl}}}{\Gamma_0} = &\frac{1}{2 m^2 c^2} \int \dd \bm{p} \  p^2 \left( \left| \psi_\text{sup}  (\bm{p}) \right|^2 - P_{\rm cl}(\bm{p}) \right) \nonumber \\
= & \  \gamma_{\rm Q}^{-1},
\end{align}
which is equal to the quantum correction to the classical time dilation contribution given in Eq.~\eqref{gammaQ} and derived in~\cite{Smith2020}. This is a surprising result because the clock model considered here, based on the spontaneous decay of an atom, observes the same quantum time dilation effect as the quantum clock considered in~\cite{Smith2020} and described in Sec.~\ref{Quantum time dilation}. This observation supports the conjecture that quantum time dilation for constant velocities is universal, affecting all clocks in the same manner.

The difference $\gamma_{\rm Q}^{-1}$ in transition rate between a coherent superposition and classical mixture of momentum wave packets can be of positive or negative sign, depending on the relative phase between two wave packets $\phi$ and their relative weight $\theta$ (see Fig.~\ref{fig1}).
For instance, for an equally weighted superposition it is seen that Eq.~\eqref{corrections} does not depend on the sum of the wave packets' momenta; it is positive for a relative phase smaller than $\phi = \pi / 2$ and becomes negative for a larger value.
In this case, the structure of the quantum contribution exhibits a distinctive peak for a given relative phase $\phi$.
If $\phi=0$, this peak occurs at a finite momentum difference, $\bar{p}_2 - \bar{p}_1 \approx 2 \Delta$; however, if the relative phase is $\phi = \pi$, the position of the peak shifts towards $\bar{p}_2 - \bar{p}_1 \approx 0$.

This behavior can be understood by analyzing the structure of the wave packets in momentum space\,---\,when the wave packets almost fully overlap, their relative phase plays a crucial role.
If the separation in momentum space between the wave packets vanishes, then there is no distinction between the coherent superposition and incoherent classical mixture, as two wave packets are identical.
As the phase approaches $\pi$, the real part of the center-of-mass wave function goes to $0$, pronouncing the imaginary part which is an antisymmetric function. This is in stark contrast to the classical mixture for which the density stays single peaked.

Surprisingly, an equally weighted superposition is not optimal for maximizing the effect of quantum time dilation, as it saturates at $-\Delta^2/2m^2c^2$ for a relative phase equal to $\phi = \pi$. The global maximum is also achieved for $\phi = \pi$, however for a slightly unbalanced superposition, $\theta \approx \pi /4 \pm (\bar{p}_2-\bar{p}_1) / 2 \sqrt{2} \Delta $.
If one considers wave packets' with average momenta much larger than their spreads, \mbox{$(\bar{p}_1+\bar{p}_2)/\Delta \gg 1$}, then this maximum becomes proportional to the sum of the momenta, $\pm \sqrt{2} \Delta |\bar{p}_1+\bar{p}_2|/4 m^2 c^2$.
This indicates that the effect of quantum coherence on the emission rate increases as the average momenta of the wave packets increases.

Note that the quantum correction $\gamma_{\text{Q}}^{-1}$ to the time dilation observed by the atom's decay rate is second order in the atom's average momentum, see Eqs.~\eqref{gammaQ} and \eqref{corrections}, analogous to the a classical time dilation contribution governed by $\gamma_{\text{C}}^{-1}$. However, linear effects, such as a Doppler shift, can also be affected by momentum coherence. Such effects can be characterized by the difference in the first moments of the momentum distributions associated with a coherent superposition and incoherent classical mixture:
\begin{align}\label{correctionslin}
\delta_{\rm Q} &\equiv \frac{1}{m c} \int \dd \bm{p} \  p \left( \left| \psi_\text{sup}  (\bm{p}) \right|^2 -  P_{\rm cl}  (\bm{p})  \right) \nonumber \\
& = \frac{\cos \phi \ \sin 4 \theta \left(\bar{p}_2-\bar{p}_1 \right) }{4 m c \left[\cos \phi \sin 2 \theta +e^{\frac{\left( \bar{p}_2 - \bar{p}_1 \right)^2}{4 \Delta^2}} \right] }.
\end{align}
The behavior of $\delta_{\rm Q}$ is qualitatively different than  $\gamma_{\rm Q}^{-1} $.
{Most notably, $\delta_{\rm Q}$ vanishes if the superposition of symmetric wave packets is equally weighted. 
On the other hand, as it is linear in momentum, it is easier to measure as the absolute magnitude of this effect is necessarily larger than the second order quantum time dilation effect characterized by $\gamma_{\rm Q}^{-1}$.
The detailed analysis of $\delta_{\rm Q}$ is provided in the Appendix~\ref{signature_sec}}.

If one considers an angular distribution of emitted photons from the decaying atom,~\eqref{fulang}, the difference between coherent and incoherent cases is given by 
\begin{align}\label{angdiff}
\frac{\Gamma_{\text{sup}} (\Theta,\Phi) \!-\Gamma_{\text{cl}} (\Theta,\Phi) }{\Gamma_0} = \Xi_1 (\Theta,\Phi) \ \delta_{\rm Q} + \Xi_2 (\Theta,\Phi) \ \gamma_{\rm Q}^{-1}.
\end{align}
Note that the angular distribution of the radiation can be affected linearly in momentum and analysis of Fig.~\ref{fig0}(b) shows that the contribution stemming from the momentum coherence is the most pronounced for photons emitted in the direction of atom's motion.
However, as this quantum correction is proportional to $\delta_{\rm Q}$, it will vanish if the atom's center-of-mass is prepared in an equally weighted superposition of momentum wave packets.
On the other hand, the amount of photons emitted perpendicular to both the dipole moment and the direction of motion is not affected linearly in atom's momentum.
Additionally, as shown in Fig.~\ref{fig0}(c), the second order contribution is the largest in this direction, suggesting that photon detection in this direction is optimal for measuring quantum time dilation.

\begin{figure}[h!tbp]
	\includegraphics[width=0.84\linewidth]{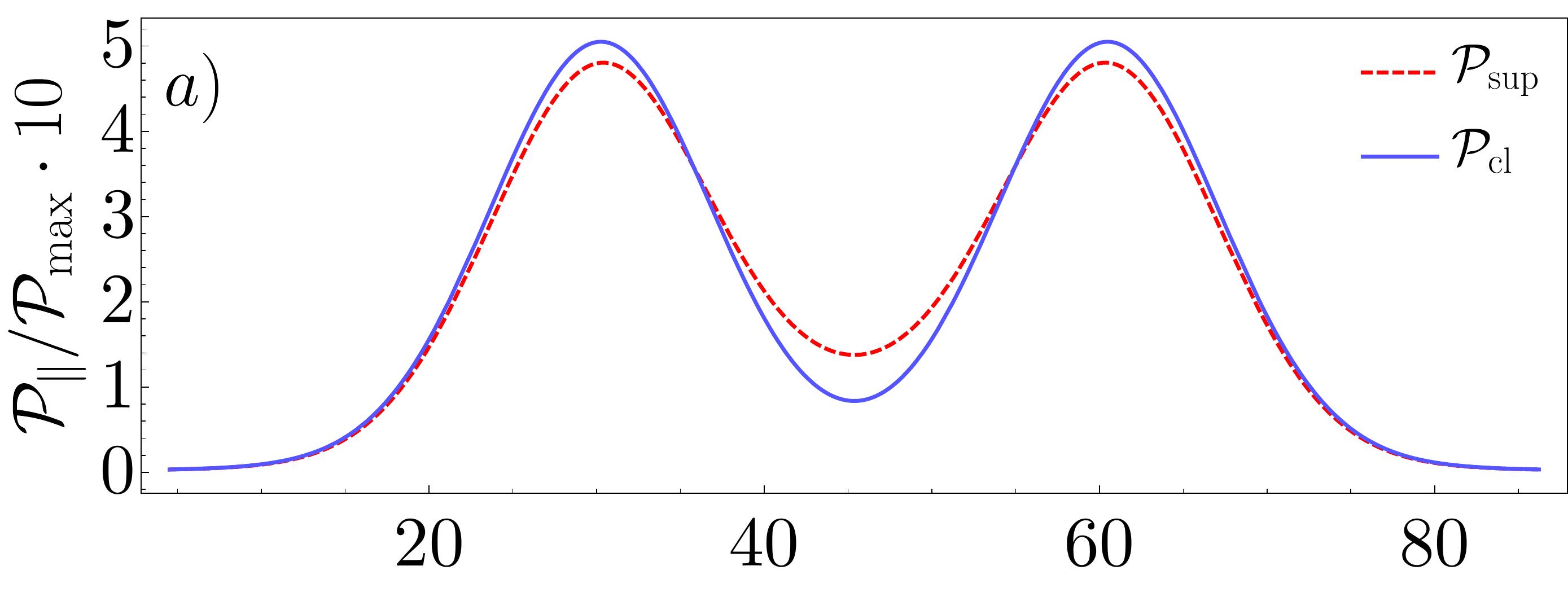}
	\includegraphics[width=0.84\linewidth]{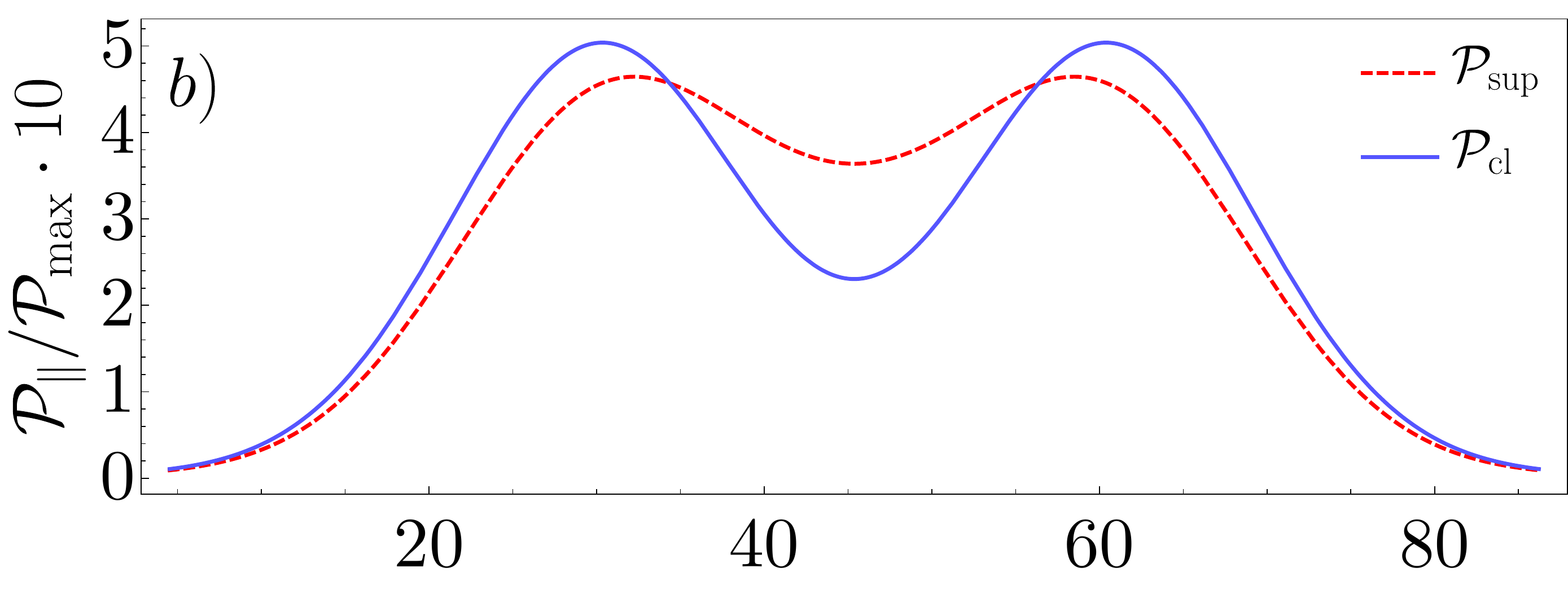}
	\includegraphics[width=0.84\linewidth]{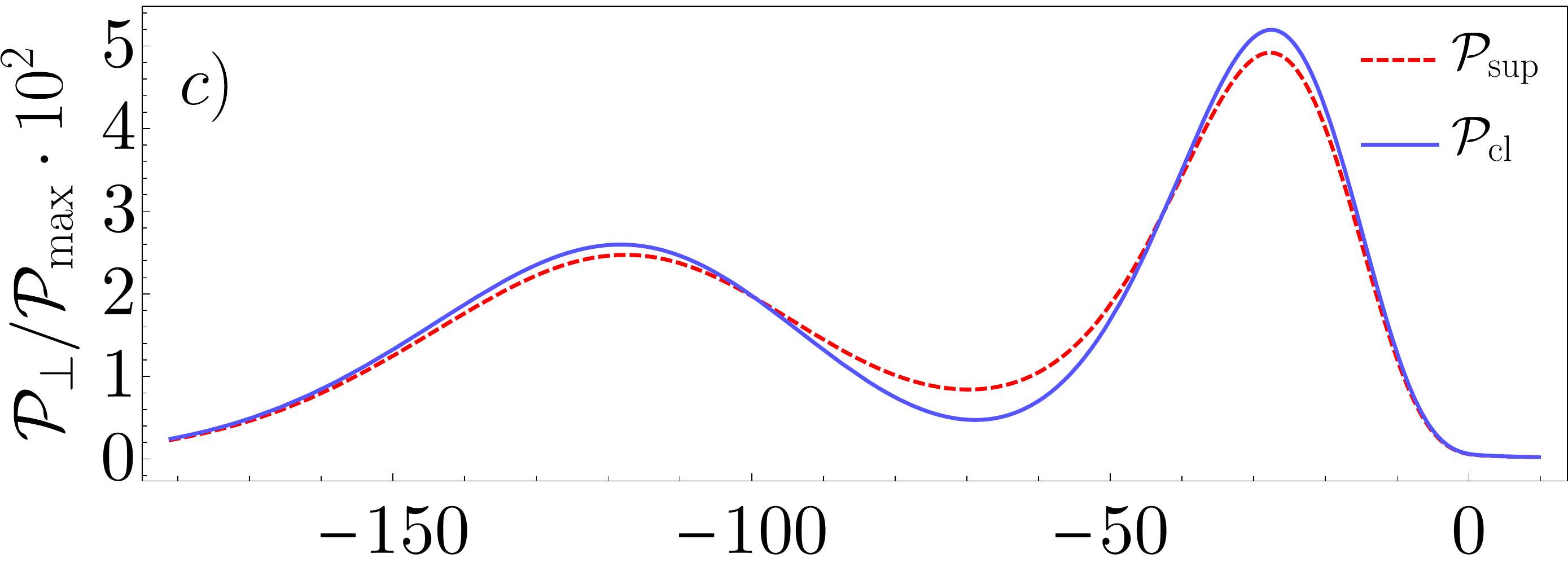}
	\includegraphics[width=0.84\linewidth]{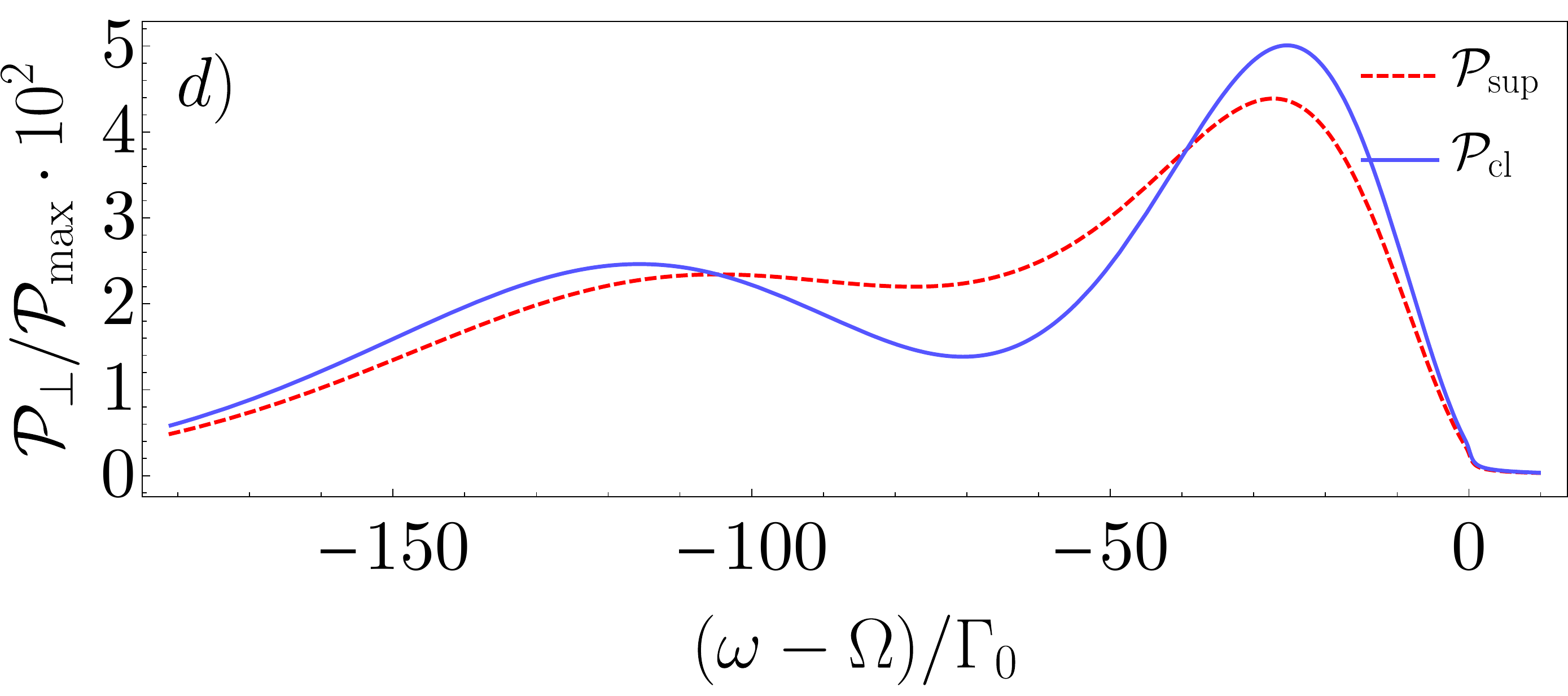}
	\caption{\label{plot2} 
		Emission line shape $\mathcal{P}(\omega)$ of the spontaneous decay of an atom that is initially prepared in a coherent superposition ($\mathcal{P}_{\text{sup}}$) and in a classical mixture ($\mathcal{P}_{\text{cl}}$) of two momentum wave packets sharply peaked at different momenta, $\bar{p}_1 = 2 \cdot 10^{-8} m c$ and $\bar{p}_2= 4 \cdot 10^{-8} m c$ {(velocities achievable for ion clocks~\cite{Brewer2019} or momentum cat states~\cite{Johnson2017})}.
		{The emission line is measured parallel, $\mathcal{P}_{\parallel}(\omega)$ or perpendicular, $\mathcal{P}_{\perp}(\omega)$ to the motion of the wave packets and is normalized to the maximum probability for a single stationary wave packet in a given case, $\mathcal{P}_{\text{max}}$.}
		In the former case, the dominant shift of the transition peak comes from the Doppler shift, while for the latter case---from the time dilation.
		{Note that transverse emission in suppressed compared to parallel emission.} 
		Panels a) and b) are calculated for a broad transition, $\Omega / \Gamma_0 \approx 1.5 \cdot 10^9$ {(e.g.\ hydrogen $^2P-{^1S}$ transition)}, while panels c) and d) are associated with the extremely narrow $\Omega / \Gamma_0 \approx 1.5 \cdot 10^{17} $ {(e.g.\ aluminium $^1S_0-{^3P}_0$ transition)}.
		It showcases the fact that quantum relativistic effects can be probed even for broad transitions, if the Doppler shift is affected.
		If the spread of the momentum wave packets is much smaller than their separation, $\Delta \ll |\bar{p}_2 - \bar{p}_1|$, coherent and incoherent cases are almost indistinguishable, with two sharp shifted peaks clearly visible (panels (a) and (c), $\Delta / mc = 6 \cdot 10^{-9}$).
		Note the broadening of structures due to a  finite spread of momentum (i.e., a homogeneous Doppler effect).
		If the momentum spread becomes larger and the overlap of the two wave packets increases (panels (b) and (d), $\Delta / mc = 8 \cdot 10^{-9}$), interference effects become visible, manifesting direct confirmation of quantum relativistic effects in the atomic spectrum.
}
\end{figure}

Similar to the total emission rate and the angular distribution, the shape of a transition line is also affected by the nonclassicality of the center-of-mass state as shown in Eqs.~\eqref{tranal0} and~\eqref{tranal01}. 
That is, the emission spectrum of an atom in a coherent momentum superposition is distinct from that of an incoherent classical mixture.
As suggested by analysis of the angular distribution of radiation, we will focus on two cases: photons emitted parallel and perpendicular to the atom's motion.
Experimentally, both scenarios can be realized by emission and absorption spectroscopy, with the latter producing an absorption line shape that can be derived from the emission shape via the Einstein coefficients. To keep the discussion simpler, we will discuss only the emission line.

First, photons emitted in the direction of motion are affected by the classical Doppler effect, shifting the center of the transition line linearly in $p/mc$, contrary to relativistic effects, that cause shifts quadratic in $p/mc$.
Analogously to quantum time dilation, the correction coming from momentum coherence to the Doppler shift can be dubbed a \textit{quantum Doppler shift}.
It is important to note that this effect only modifies the shape of the emission spectrum, not the total emission rate, which is affected by quantum time dilation.
The quantum Doppler effect smooths the contrast between two transition rate peaks associated with two different Doppler shifted emission lines; see Fig.~\ref{plot2}(a)-(b). The difference between the quantum and classical Doppler shift is most pronounced in between the emission peaks, which may suggest that the postselection of the final momentum of the atom may further enhance the effect.
A quantitative analysis describing how the quantum Doppler effect modifies emission line shape is provided in the Appendices~\ref{signature_sec} and \ref{emission_app}.

In the case of an emission perpendicular to the direction of motion, the classical Doppler shift is not present; that is, corrections linear in $p/mc$ are not present.
The center of the transition line is shifted quadratically in $p/mc$, heralding the onset of relativistic effects.
Relativistic corrections of this magnitude can be measured in state-of-the-art experiments~\cite{chouOpticalClocksRelativity2010} and are affected by a momentum coherence analogously to the quantum Doppler shift; see Fig.~\ref{plot2}(c)-(d).

\section{Experimental considerations}

Since the initial prediction of wave-particle duality by de Broglie, atomic interferometry has witnessed tremendous conceptual and technological progress~\cite{Cronin2009}.
With the advent of large momentum beamsplitters~\cite{Clade2009,Chiow2011,McDonald2013,Rudolph2020}, superpositions of atomic beams travelling along distinct trajectories have been realized, leading to quantum-based alternatives to classical gravimeters, gradiometers, and accelerometers~\cite{Kasevich1991,Hu2013,Dutta2016,Abend2016,rouraGravitationalRedshiftQuantumClock2020a,geiger2020highaccuracy}. In these settings, the usual strategy is to suppress radiation losses as they disrupt phase relations between arms of an interferometer~\cite{Rudolph2020}. 

In contrast, our proposal is to test the nonclassicality of center-of-mass motion through spectroscopic measurements. As shown above, a coherent superposition of relativistic momenta affects the spontaneous emission rate beyond classical time dilation effects, thus offering a spectroscopic signature of quantum time dilation. Moreover, spectroscopic methods offer a plethora of schemes that might witness similar quantum-relativistic effects\,---\,among others, stimulated absorption and emission spectroscopy, and techniques involving M\"ossbauer effect and Rydberg states~\cite{Cronin2009}. In particular atomic clocks provide a natural test bed for a relativistic theory due to their unparalleled accuracy~\cite{Hinkley2013, Ludlow2015a, Brewer2019}. Such clocks have been used to observe classical time dilation at velocities as low as several meters per second~\cite{chouOpticalClocksRelativity2010}.

To go beyond and measure quantum time dilation, one has to deal with experimental challenges that involve an interplay between different time scales. The lifetime of the excited atom has to be long enough to allow for the creation of sufficient momentum separation and to precisely excite the atomic beam. However, the lifetime cannot be longer than the coherence time of the center-of-mass superposition. Fortunately, due to advanced methods of phase imprinting in atomic systems~\cite{Dobrek1999,Denschlag2000}, initial states maximizing the quantum contribution might be engineered.

Specifically, there are promising experimental setups offering access to the accuracy needed to observe quantum-relativistic effects. Among others, quantum clocks based on aluminium ions have recently achieved precision going beyond leading relativistic corrections~\cite{chouOpticalClocksRelativity2010,Ludlow2015a,Brewer2019}.
In such setups, an aluminium ion is confined to a quadrupole trap acting with an effective harmonic potential and is prepared close to zero-point motion energy by advanced cooling techniques.
The ion is perturbed triggering oscillatory motion in a coherent-state-like fashion.
Spectroscopic measurement allows for resolving the resulting frequency shift due to the ion's motion below $10^{-18} \Omega$, which is far below the leading relativistic correction of $10^{-15} \Omega$~\cite{Brewer2019}.

To observe quantum time dilation. a coherent momentum superposition of such an ion must be prepared, a momentum Schr\"odinger cat state. 
This on its own is a state-of-the-art task, however recent advances have reported tremendous progress in this direction~\cite{Kasevich1991,Lo2015,Kovachy2015,Johnson2017}.
For example, ytterbium ions have been prepared in mesoscopic superpositions of motional states~\cite{Johnson2017}. 

Creation of such an ion that exhibits a narrow transition line is required for observation of the quantum time dilation effect in order to resolve the associated frequency shift which is second order in the average momenta of the wave packets. Mean velocity of a trapped ion easily resolvable for an ion clock in a laboratory, \SI{5}{m/s}, corresponds to a coherent state $\ket{\alpha}$ with $\abs{\alpha} \approx 12$.  State-of-the-art separation between coherent states can go up to $\abs{\alpha} \approx 24$~\cite{Johnson2017}, showing that a coherent superposition of momenta can be achieved within spectroscopic resolution.

Generally speaking, in atomic clock experiments a transition line is measured.
The difference in transition line shape exhibited by a coherent superposition and incoherent classical mixture of momentum wave packets, as shown in Figs.~\ref{plot2}(c)-(d), would be confirmation of quantum time dilation.
The difference between these two cases is most pronounced for a frequency that corresponds to the average of mean momenta of the superimposed wave packets.
The upshift of the transition probability due to momentum coherence at this specific point be as much as 40$\%$ if the parameters of the superposition are optimized, while also not being far away from resonance as depicted in Fig.~\ref{plot2}(d).
Changes of this magnitude are routinely measured in state-of-the-art experiment involving ion clocks~\cite{chouOpticalClocksRelativity2010,Brewer2019}.

In such an experimental setup the balance between the ability to create a superposition of momentum wave packets and the precision of a given ion clock is crucial in order to observe a signature of quantum time dilation. Other obstacles exist, such as excess motion, secular motion, the quadratic Zeeman effect, deviations from harmonic trapping etc.~\cite{Ludlow2015a}, which are usually well resolved in atomic clock experiments. Nonetheless, additional work needs to be done to characterize these effects in presence of relativistic momentum coherence to deduce the optimal experimental scenario.

On the other hand, experiments involving large momentum transfer between between light and atomic beams might also provide a viable alternative for a measurement of quantum time dilation~\cite{McGuirk2000,Chiow2011,Graham2013,Plotkin-Swing2018,Rudolph2020}, where limitations due to excited state decay would work as an advantage.
As they are not yet realized for narrow transitions at the level of atomic clocks, such an upgrade is widely sought as a key ingredient for gravitational wave~\cite{Graham2013} and dark matter detectors~\cite{Kennedy2020}.
Advances in this direction could make possible a measurement of quantum time dilation in such a setup, as narrower transitions would enhance the spectroscopic precision.

The momentum separation currently achieved in these experiments is around $140 \  \hslash k$ for the strontium transition $^1S_0$-$^3P_1$, where $k$ is the magnitude of the wave vector of the incident light~\cite{Rudolph2020}.
Such a momentum separation corresponds to a \SI{1}{m/s} velocity difference between two clouds of atoms. Larger momentum transfer is expected in the future, with experimental proposals promising up to $1000 \  \hslash k$~\cite{Rudolph2020}. 
The widths of these momentum wave packets are relatively small for the detection of quantum time dilation, with a rms Doppler width of 25 kHz, corresponding to a velocity width of \SI{0.02}{m/s}~\cite{Rudolph2020}.
As the maximal value of quantum time dilation scales like $\frac{\Delta}{mc} \frac{p_2-p_1}{m c} \Gamma_0$, it is still too far away to be measured in these experiments.
However, as the interest in large momentum transfer grows with a potential use in low energy studies of quantum gravitational effects, the engineering of a quantum time dilation experiment might be possible in the near future. Such an experiment could be achieved by either a larger momentum spread of a single wave packet or a larger momentum separation between momentum wave packets.

Similar to the superposition of momentum wave packets considered above, quantum effects manifest for atoms in spatial superpositions~\cite{Rzazewski1992,Steuernagel1995,Saba2005,Fedorov2005,stritzelbergerCoherentDelocalizationLightmatter2020}. From an experimental point of view, such studies have been helpful in analyzing phase coherence in Bose-Einstein condensates interacting with light~\cite{Saba2005}. These systems provide an extremely clean environment to study atomic systems with possibility of fine tuning of interactions and spatial geometry. As such, they might accommodate experiments with coherent superposition of momentum wave packets (e.g.\ non-equilibrium dynamics in double-well trap~\cite{Thomas2002}).
Moreover, in contrast to ion clocks, experiments involving large momentum transfer or trapped ultracold gases have still not achieved the necessary velocities to be sensitive to relativistic motion of particles. However, such experiments have the advantage that they utilize large ensembles of atoms, which results in a stronger signal that should scale proportionally to the particle number.

\section{Conclusions and outlook}

We have proposed a spectroscopic signature of quantum time dilation that manifests in the spontaneous emission rate (lifetime) of an excited atom moving in a superposition of relativistic momentum wave packets. We have shown that the total transition rate is strongly affected by momentum coherence in the center-of-mass state of the atom. Furthermore, the quantum contribution to the time dilation observed by an excited atom can be either positive or negative, depending on the relative phase between the superposed momentum states and is within reach of the existing experimental setups~\cite{chouOpticalClocksRelativity2010,Clade2009,Chiow2011,McDonald2013,Rudolph2020}. We observed that the quantum contribution to the time dilation of the atomic lifetime in Eq.~\eqref{corrections} was equal to the quantum time dilation observed on average by the ideal clock considered in ~\cite{Smith2020,smithQuantumTimeDilation2020}. These two clocks are built on very different mechanisms (spontaneous decay and a particle on a line), so this result indicates that quantum time dilation is universal, affecting all clocks in the same way.

The effects of quantum time dilation on atomic spectra complement the growing literature on relativistic clock interferometry, which also probes the effects of suppositions of clocks experiencing different proper times due to both special or general relativistic effects~\cite{zychQuantumInterferometricVisibility2011a,bushevSingleElectronRelativistic2016,zychGravitationalMassComposite2019a,lorianiInterferenceClocksQuantum2019a,khandelwalGeneralRelativisticTime2019,rouraGravitationalRedshiftQuantumClock2020a,geiger2020highaccuracy}. In contrast, we propose a spectroscopic signature of proper time superpositions that can probe quantum theory and relativity in the regime in which coherence across relativistic momentum wave packets plays a role.

We have also characterized a quantum Doppler effect that occurs when the center-of-mass of an atom is in a superposition of momentum wave packets. The effect is present in the shape of the emission spectrum, affecting its structure by smoothing the contrast between distinctive Doppler-shifted peaks. \\

\textit{Note added.}---Following the initial posting of this article a related preprint on delocalized center-of-mass atomic wave functions and the light-matter interaction appeared~\cite{lopp2020quantum}.

\begin{acknowledgments}
We would like to thank Kazimierz Rz\k{a}{\.z}ewski for fruitful discussions and pointing out~\cite{Rzazewski1992}, and Mehdi Ahmadi for his valuable contributions to the ideas developed here.
This work was supported by the Natural Sciences and Engineering Research Council of Canada and the Dartmouth Society of Fellows.
This article has been supported by the Polish National Agency for Academic Exchange under Grant No. PPI/PZA/2019/1/00094/U/00001.
\end{acknowledgments}

\appendix
\section{Momentum wave packets and signatures of coherence}\label{signature_sec}
\begin{figure*}[t]
	\includegraphics[height=0.31\linewidth]{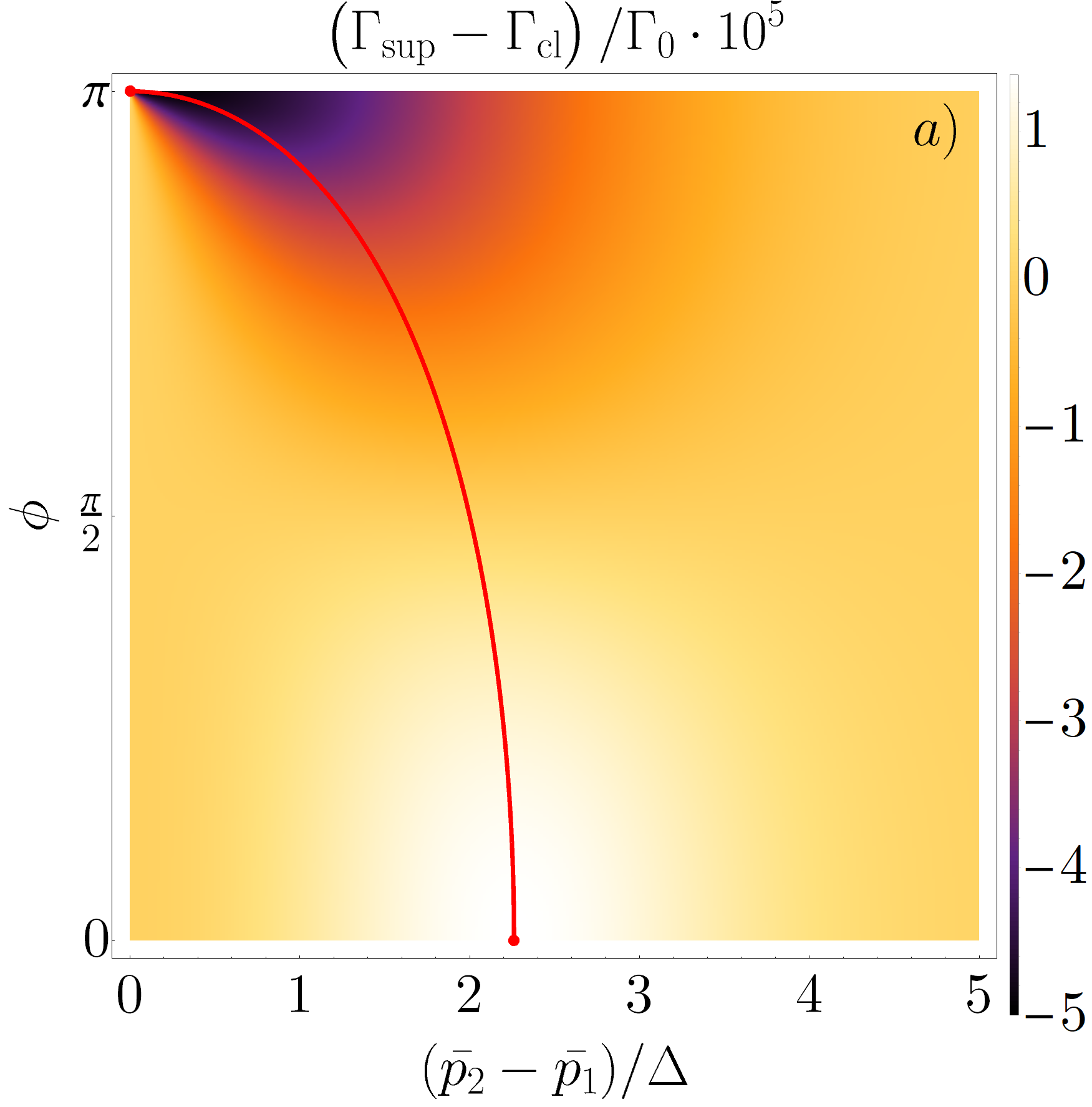} \quad 
	\includegraphics[height=0.31\linewidth]{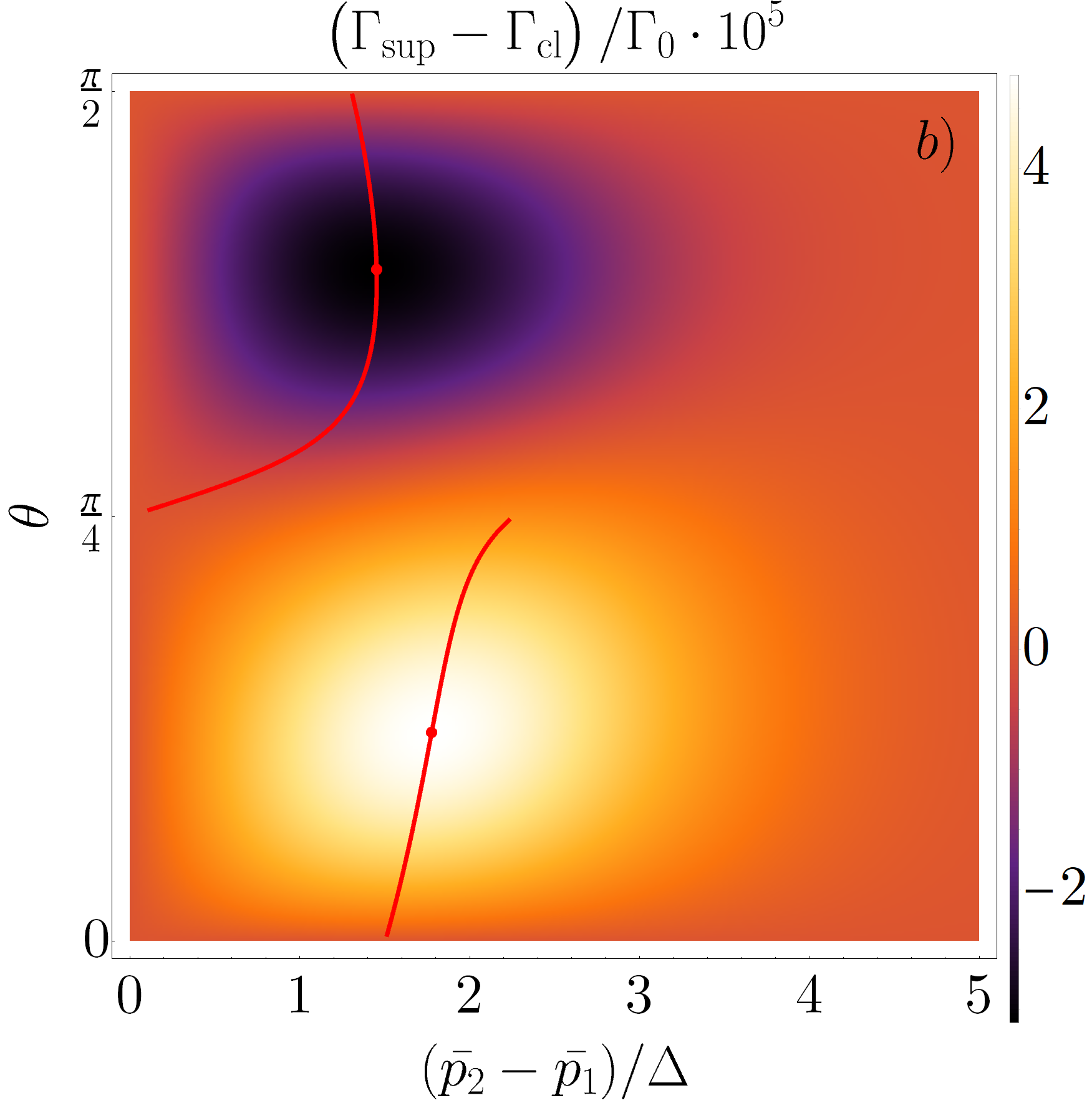} \quad 
	\includegraphics[height=0.31\linewidth]{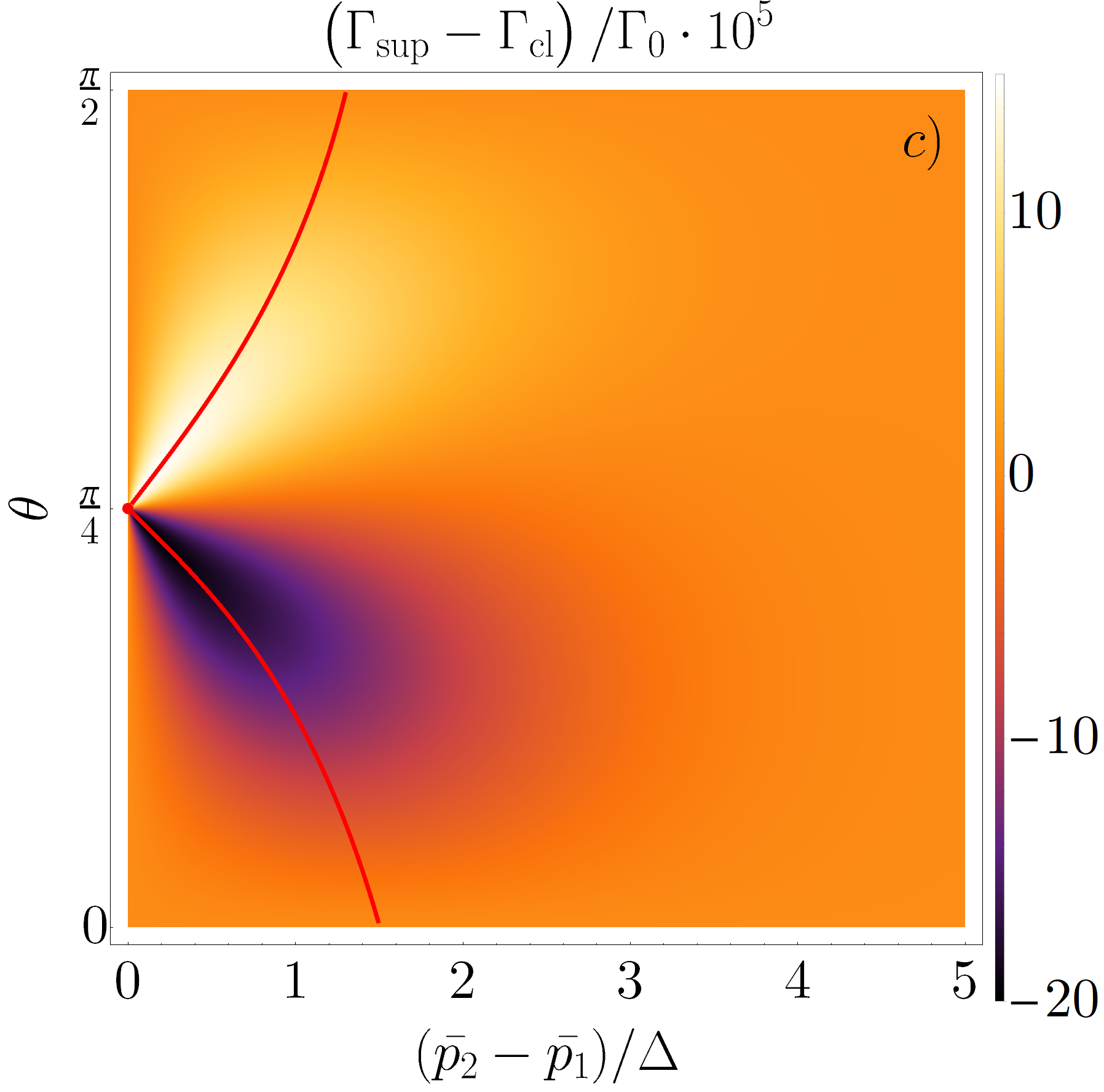}
	\caption{\label{figsup1} 
	Fig.~\ref{fig1} from the main text; shown in Appendix for a purpose of comparison with Fig.~\ref{figsup2}. The difference in total emission rates between a superposition and a classical mixture of two momentum wave packets of an atom as a function of the wave packets' momentum difference and their relative phase and weight.
	The red line marks the maximum value of the effect for a given relative phase or relative weight, while the red circles signify maximum and minimum values across the given subplot.
	a) equal weighted superposition of momentum wave packets, $\theta = \pi / 4$.
	b) Relative phase fixed at $\phi = 0$.
	The red line is not continuous at the point $\theta=\pi/4$, and it sharply ends at $(\bar{p}_2-\bar{p}_1)/\Delta=2\sqrt{1+W_0(1/e)}\approx 2.261$, where $W_0$ is the principal branch of the Lambert $W$ function. 
	c) Relative phase fixed at $\phi = \pi$.
	It can be seen that two extrema exist for small values of $(\bar{p}_1-\bar{p}_2)/\Delta$ and $\theta\approx\pi/4$. One can show that these extrema are placed at
$\theta=\frac{\pi}{4}-\frac{1}{4}\frac{ \bar{p}_2-\bar{p}_1}{ \bar{p}_2+\bar{p}_1} (1\pm\sqrt{1+2 (\bar{p}_2+\bar{p}_1)^2/\Delta^2})$ and their corresponding values are $\pm\Delta^2(\sqrt{1+2(\bar{p}_1+\bar{p}_2)^2/\Delta^2}-1 )/4m^2c^2$. 
	Nonzero value for a finite momentum difference signifies the phenomenon of quantum time dilation.
	In each of the panels, the momentum spread of each of wave packets is $\Delta = 0.01 m c$ and the sum of their average momenta is equal to $\bar{p}_1 + \bar{p}_2 = 0.05 m c$.
}
\end{figure*}

\begin{figure*}[t]
	\includegraphics[height=0.31\linewidth]{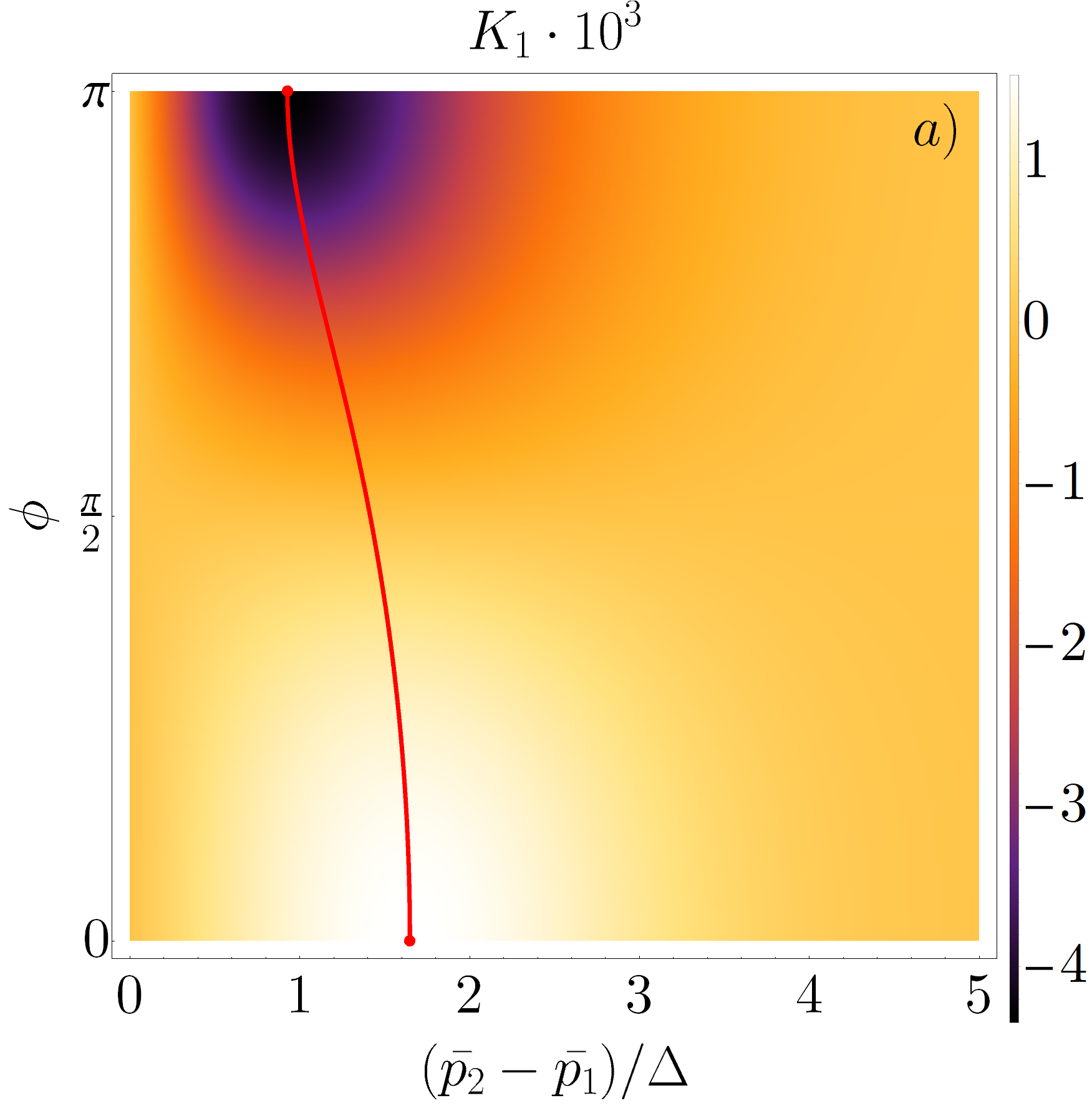} \quad 
	\includegraphics[height=0.31\linewidth]{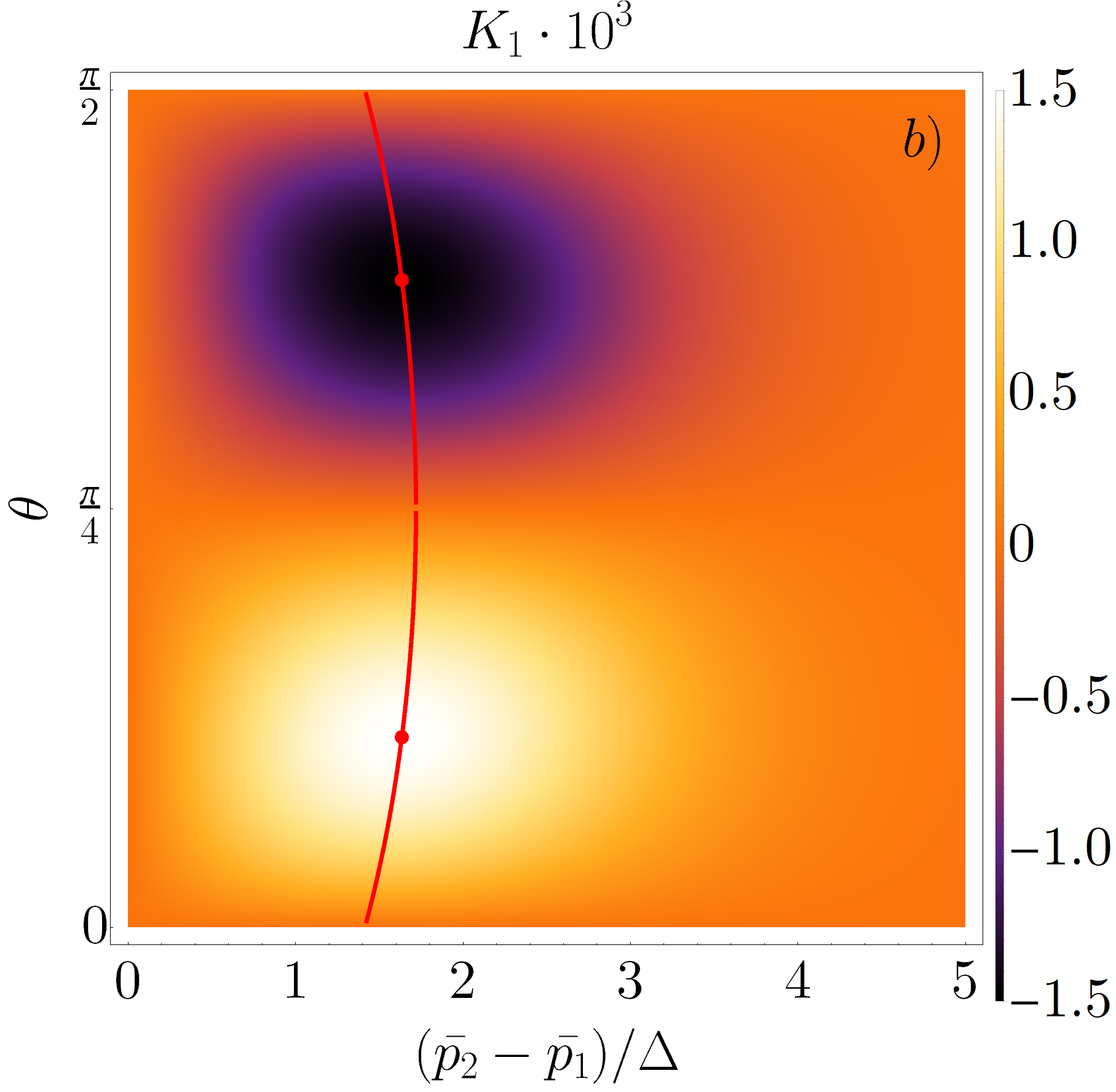} \quad 
	\includegraphics[height=0.31\linewidth]{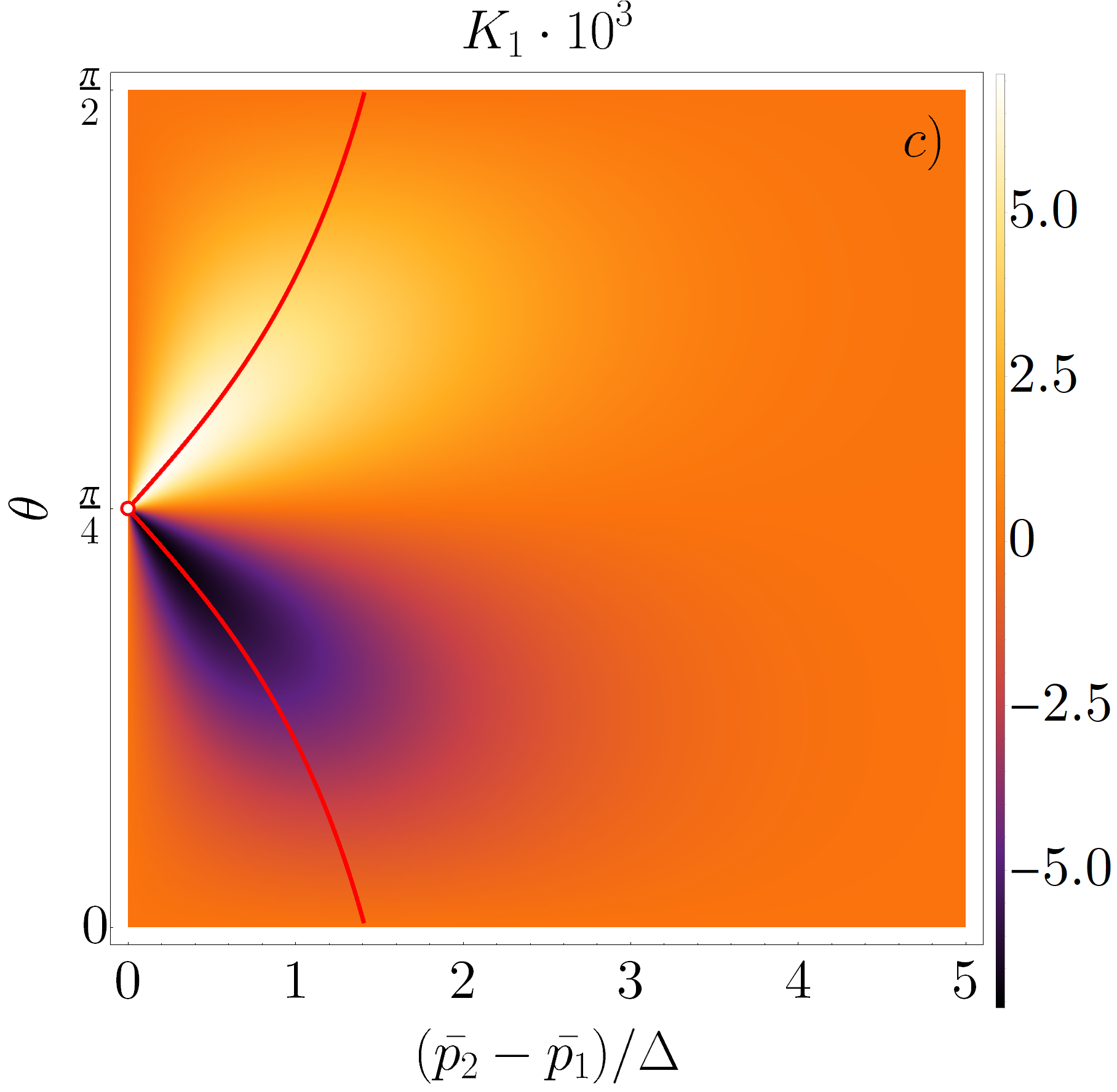}
	\caption{\label{figsup2} 
	The difference in first moments $K_1 = \delta_{\rm Q}$ of the momentum distributions associated with a superposition and a classical mixture of two momentum wave packets as a function of the wave packets' momentum difference and their relative phase and weight, which quantifies the difference in magnitude of angular distribution of emission: a) unequal weighted superposition of momentum wave packets, $\theta = \pi / 8$, b) Relative phase fixed at $\phi = 0$, c) Relative phase fixed at $\phi = \pi$.
	The red line marks the maximum value of the effect for a given relative phase or a relative weight, while the red circles signify maximum and minimum values across the whole plot.
	Nonzero values for a finite momentum difference signifies the phenomenon of quantum time dilation.
	In each of the panels, the momentum spread of each of the wave packets is $\Delta = 0.01 m c$ and the sum of their average momenta is equal to $\bar{p}_1 + \bar{p}_2 = 0.05 m c$.
}
\end{figure*}

Here, explicit forms of momentum wave packets from the main text are presented.
It is assumed that the atom in consideration moves along the $z$-direction with its momentum distribution in perpendicular directions well localized around $p_x=p_y=0$.
Moreover, we consider the atom to be either in a coherent superposition of two Gaussian wave packets:
\begin{equation}
    \psi_{\text{sup}}(p) = \mathcal{N}\left[ \cos \theta e^{-\frac{(p-\bar{p}_1)^2}{2\Delta^2}}
+  e^{i\phi} \sin \theta e^{-\frac{(p-\bar{p}_2)^2}{2\Delta^2}} \right]\!, 
\end{equation}
where $\mathcal{N} = [ \sqrt{\pi}  \Delta ( 1+ \cos \phi \sin 2 \theta e^{-(\bar{p}_1-\bar{p}_2)^2/4\Delta^2 }) ]^{-1/2}$, or in an incoherent mixture described by the momentum distribution:
\begin{align}
P_{\text{cl}}=\frac{1}{\sqrt{\pi} \Delta} \left[ \cos^2 \theta e^{-\frac{(p-\bar{p}_1)^2}{\Delta^2}} + \sin^2 \theta e^{-\frac{(p-\bar{p}_2)^2}{\Delta^2}} \right].
\end{align}
The difference between coherent superpositions and classical mixtures of momentum wave packets can be characterized by the difference in moments associated with their respective momenta distributions:
\begin{align}
K_j \equiv \frac{1}{j! m^j c^j} \int \dd p \  p^j \left[ \abs{\psi_{\text{sup}} \left(p\right)}^2 - P_{\rm cl} \left(p\right) \right].
\end{align}
In case of Gaussian wave packets considered above, $K_1$ and $K_2$ take the explicit form:
\begin{align}
K_1&= \frac{\cos \phi \sin 4 \theta \left(\bar{p}_2-\bar{p}_1 \right) }{4 m c \left[\cos \phi \sin 2 \theta +e^{\frac{\left( {\bar{p}_2} - {\bar{p}_1} \right)^2}{4 \Delta^2}} \right] } = \delta_{\rm Q}, \nonumber \\
K_2&=  \frac{ \cos \phi \sin 2 \theta \! \left[ \left(\bar{p}_2-\bar{p}_1 \right)^2-2 \!\left(\bar{p}_2^2-\bar{p}_1^2\right) \! \cos 2 \theta \right] }{8 m^2 c^2 \left[\cos \phi \sin 2 \theta +e^{\frac{\left( \bar{p}_2 - \bar{p}_1 \right)^2}{4 \Delta^2}} \right]} = \gamma_{\rm Q}^{-1}.
\end{align}
Analysis of $K_2$ is presented in Fig.~\ref{figsup1} and of $K_1$ in Fig.~\ref{figsup2}.

Note that $K_1$ vanishes for an equally weighted superposition $\theta=\pi/4$.
Let us show that this feature is a common feature of all symmetric wave packets.
Let $\varphi(p)$ be a normalized wave packet symmetric with respect to $p=0$.
Then, we can write an equally weighted, coherent superposition of two momentum wave packets as 
\begin{equation}
    \varphi_{\text{sup}}(p) = \frac{\mathcal{N}}{\sqrt{2}}\left[ \varphi(p-p_1)
+  e^{i\phi} \varphi(p-p_2) \right]\!, 
\end{equation}
with $\mathcal{N}^2 = \left[1+\cos \phi \int \dd p \ \varphi(p-p_1) \varphi(p-p_2) \right]^{-1}$.
For the corresponding classical mixture, the momentum distribution takes the form
\begin{align}
P_{\text{cl}}=\frac{1}{2} \left( \varphi^2(p-p_1) + \varphi^2(p-p_2) \right).
\end{align}
Then, by an explicit evaluation, one finds that
\begin{align}\label{k1ex}
\frac{2 m c}{\mathcal{N}^2} K_1  &= \frac{2 }{\mathcal{N}^2}  \int \dd p \  p \left[ \abs{\varphi_{\text{sup}} \left(p\right)}^2 - P_{\rm cl} \left(p\right) \right]  \nn \\
&=-\cos \phi \Big[ \left(  \int \dd p \ \varphi(p-p_1) \varphi(p-p_2)\right) \nn \\
&\quad \times \left(  \int \dd p \ p \left(\varphi^2(p-p_1)+ \varphi^2(p-p_2)\right)\right)  \nn \\
&\quad-2 \int \dd  p \ p \, \varphi(p-p_1) \varphi(p-p_2) \Big].
\end{align}
The term in the third row of Eq.~\eqref{k1ex} equals $p_1 + p_2$, because $\varphi(p-p_{1,2})$ are normalized and well localized around $p_{1,2}$.
By substituting $p \rightarrow p-(p_1+p_2)/2$ and utilizing the fact that expression $\varphi(p-p')\varphi(p+p')$ is an even function of $p$, one finds that
\begin{align}
&2 \int \dd  p \ p \ \varphi(p-p_1) \varphi(p-p_2)  \nn\\
&\hspace{.5in}= (p_1+p_2) \int \dd  p \ \varphi(p-p_1) \varphi(p-p_2).
\end{align}
Substituting these two results into Eq.~\eqref{k1ex} it is seen that that $K_1 = 0$. The only necessary condition is for $\varphi(p)$ to be an even function with respect to $p=0$.
 
\section{Derivation of the emission rate and spectrum shape}\label{emission_app}
Let us focus on the angular distribution of the emitted radiation.
The angular distribution can be obtained by omitting angular integration in Eq.~\eqref{total}.
Note that such an approach does not explicitly utilize the photon distribution that comes from an integration over the probabilities $|\beta_{\bm{k}, \xi}|^2$.
However, as shown in Ref.~\cite{Wilkens1994} this approach is consistent with the special relativity, reproducing dipole pattern of radiation in the comoving frame.
Writing explicitly Eq.~\eqref{total} gives (see Eqs. (9)-(11) in Ref.~\cite{Sonnleitner2017} for a derivation)
\begin{align} 
 \Gamma & \!=\! \lim_{t \rightarrow \infty} \frac{\dd}{\dd t} \sum_{\bm{k}, \xi} \int \dd \bm{p} \, \left| \beta_{\bm{k}, \xi} \left( \bm{p},t \right) \right|^2 \nonumber \\
 & = 2 \pi \int \dd \bm{p}  \left| \psi \left( \bm{p} \right) \right|^2 \sum_{\bm{k}, \xi}  \frac{  \omega_k}{ 2 \hbar \epsilon_0 (2 \pi c)^3} g_{\bm{k}, \xi}^2 (\bm{p})  \nonumber \\ 
 &\quad  \times \delta \left(\Omega - \omega_k + \frac{1}{m} \bm{k} \cdot \left( \bm{p} - \hbar \bm{k} / 2\right) \right).
\end{align}
By taking a continuous limit of the summation over $\bm{k}$,
\begin{align} 
\sum_k \rightarrow \int \dd \bm{\kappa} \int \dd \omega  \  \omega^2 \sin \theta,
\end{align}
where $\bm{\kappa} = \frac{\bm{k} c}{\omega} = (\sin \Theta \cos \Phi, \sin \Theta \sin \Phi, \cos \Theta)$ and $\int \dd \bm{\kappa} = \int_{0}^{\pi} \dd  \Theta \int_0^{2 \pi} \dd \Phi$, we can omit the integral over direction $\bm{\kappa}$ to get the angular distribution:
\begin{align} \label{angular}
  \Gamma (\Theta, \Phi) &= \frac{  \pi}{  \hbar \epsilon_0 (2 \pi c)^3}  \int \dd \bm{p}  \left| \psi \left( \bm{p} \right) \right|^2 \sum_{\xi} \int_0^{\infty} \dd \omega \, \omega^3 \sin \Theta\nonumber \\
  &\quad \times  g_{\bm{k}, \xi}^2 (\bm{p}) \, \delta\! \left(\Omega - \omega + \frac{1}{m} \bm{k} \cdot \left( \bm{p} - \hbar \bm{k} / 2\right) \right).
\end{align}
{Note we are now working in the continuous limit and so the sum over $\omega_k$ has been replaced with an integral over~$\omega$.}

Under the assumption that the atom is heavy the coupling constant $g_{\bm{k}, \xi}^2 (\bm{p}) $ can be expanded to first order in $\hbar \Omega / m c^2$ and to second order in $p^2 / m^2 c^2$:
\begin{align}
 g_{\bm{k}, \xi}^2 (\bm{p}) &\approx \left( \bm{d} \cdot \bm{\epsilon}_{\bm{k}, \xi} \right)^2  \nonumber \\
& \quad +\frac{2}{m c} \left( \bm{d} \cdot \bm{\epsilon}_{\bm{k}, \xi} \right)\left( \bm{p} - \frac{\hslash \omega_k \bm{\kappa}}{2 c} \right)\cdot \left[ \left( \bm{\kappa} \times \bm{\epsilon}_{\bm{k}, \xi} \right) \times \bm{d} \right]  \nonumber \\
&\quad +\frac{1}{m^2 c^2} \left( \bm{p} \cdot \left[ \left( \bm{\kappa} \times \bm{\epsilon}_{\bm{k}, \xi} \right) \times \bm{d} \right] \right)^2.
\end{align}
Making use of the vector equalities:
\begin{widetext}
\begin{align}
\sum_{\xi} \left( \bm{d} \cdot \bm{\epsilon}_{\bm{k}, \xi} \right)^2 &= d^2 - \left(\bm{d} \cdot \bm{\kappa} \right)^2, \\
\sum_{\xi} \left( \bm{d} \cdot \bm{\epsilon}_{\bm{k}, \xi} \right) \bm{A}\cdot \left[ \left( \bm{\kappa} \times \bm{\epsilon}_{\bm{k}, \xi} \right) \times \bm{d} \right] &= \left( \bm{d} \cdot \bm{\kappa} \right) \left( \bm{A} \cdot \bm{d} \right) -d^2 \left( \bm{A} \cdot \bm{\kappa} \right), \\
\sum_{\xi} \left( \bm{A} \cdot \left[ \left( \bm{\kappa} \times \bm{\epsilon}_{\bm{k}, \xi} \right) \times \bm{d} \right] \right)^2 &= d^2 \left(\bm{A} \cdot \bm{\kappa} \right)^2 + A^2 \left(\bm{d} \cdot \bm{\kappa} \right)^2 - 2 \left(\bm{A} \cdot \bm{\kappa} \right) \left(\bm{d} \cdot \bm{\kappa} \right) \left(\bm{A} \cdot \bm{d} \right),
\end{align}
\end{widetext}
with $\bm{A} = \frac{2}{mc} \bm{p} - \frac{\hslash }{mc} \bm{k}$, it follows that
\begin{align} 
&\sum_{\xi} g_{\bm{k}, \xi}^2 (\bm{p}) 
\approx d^2 \Big[ \kappa_{\perp}^2 \left(1+\frac{\hslash \omega_k}{m c^2}\right) - \frac{2}{m c} \bm{p} \cdot \bm{\kappa}_{\perp}  + \nonumber \\
&\frac{1}{m^2 c^2} \left(\bm{p} \cdot \bm{\kappa} \right)^2 - \frac{2}{m^2 c^2} \left(\bm{p} \cdot \bm{\kappa} \right) \kappa_{\parallel} p_{\parallel} + \frac{1}{m^2 c^2} p^2 \kappa_{\parallel}^2 \Big],
\end{align}
where $\perp$ and $\parallel$ indicate projections perpendicular and parallel to the vector $\bm{d}$. 

We now go back to the angular distribution, in which we have to compute the following integral:
\begin{equation}
\label{integral}
\sum_{\xi} \int_0^{\infty}\!\dd \omega \, \omega^3 \sin \Theta \, g_{\bm{k}, \xi}^2 (\bm{p}) \, \delta\! \left(\!\Omega - \omega + \frac{1}{m} \bm{k} \cdot \left( \bm{p} - \hbar \bm{k} / 2\right) \!\right)\!.
\end{equation}
Again, supposing that the atom moves in the $z$-direction with its momentum distribution in the perpendicular directions given by delta functions centered at $p_x=p_y=0$. Thus we can consider $\bm{p} = (0,0,p)$. We will also suppose that $\bm{d}=(d,0,0)$ points in a direction perpendicular to $\bm{p}$.
Then, Eq.~\eqref{integral} simplifies to
\begin{align}
 d^2 \int_0^{\infty}\dd \omega \ \eta (\omega) \delta \left( \lambda(\omega)\right) =  d^2 \frac{\eta(\omega_0)}{|\lambda'(\omega_0)|},
\end{align}
where
\begin{align} \label{defeqs}
 \eta (\omega) &\equiv \omega^3 \sin \Theta \   \Bigg[  \!\left(1-\sin^2 \Theta \cos^2 \Phi\right) \left(1+\frac{\hslash \omega}{m c^2}\right) \nonumber \\
&\quad - \! \frac{2p }{m c}  \cos \Theta  + \frac{p^2}{m^2 c^2}  \left( \cos^2 \Theta \sin^2 \Phi + \cos^2 \Phi \right)  \! \Bigg],\nonumber \\
 \lambda(\omega) &\equiv \Omega - \omega + \omega \frac{p}{m c }  \cos \Theta - \frac{\omega^2 \hbar}{2 m c^2}, \nn \\
 \omega_0 &\equiv \Omega \left( 1+  \frac{p}{m c} \cos \Theta + \frac{p^2}{2 m^2 c^2} \cos 2 \Theta- \frac{\hbar \Omega}{2 m c^2} \right) .
\end{align}
\begin{figure*}[h!btp]
	\includegraphics[width=.31\textwidth ]{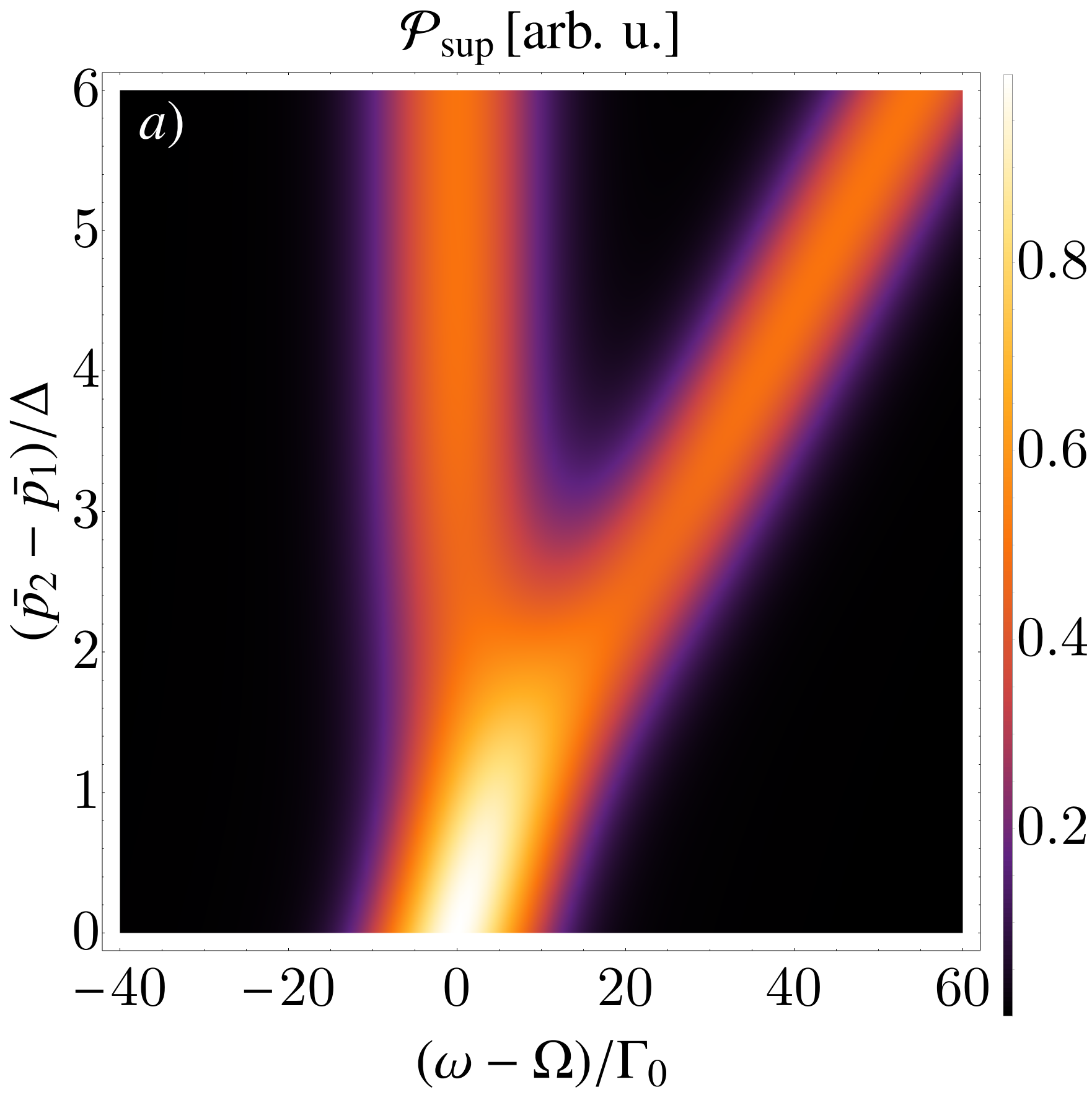} \quad 
		\includegraphics[width=.32\linewidth]{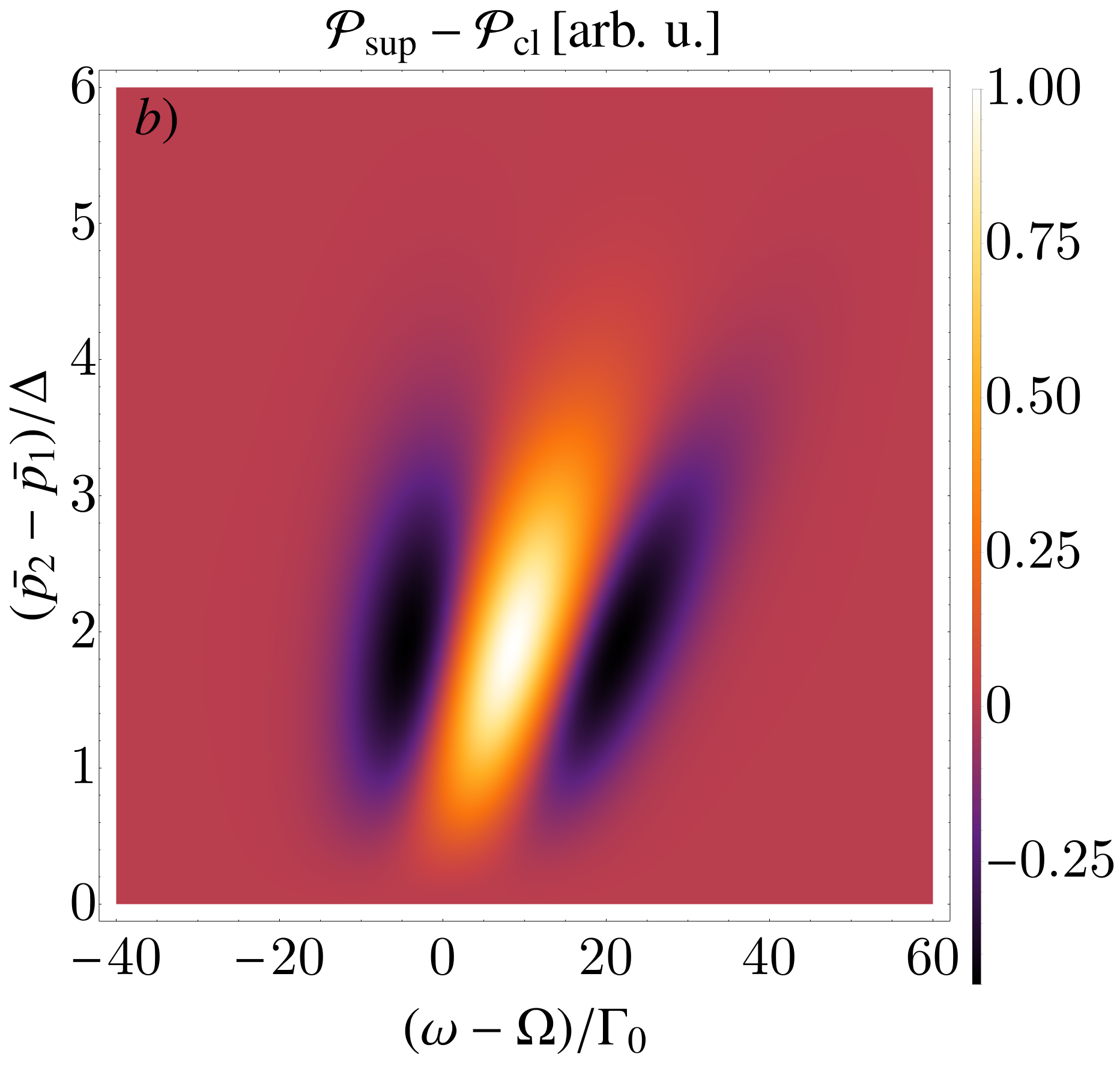} \quad 
	\includegraphics[width=.31\linewidth]{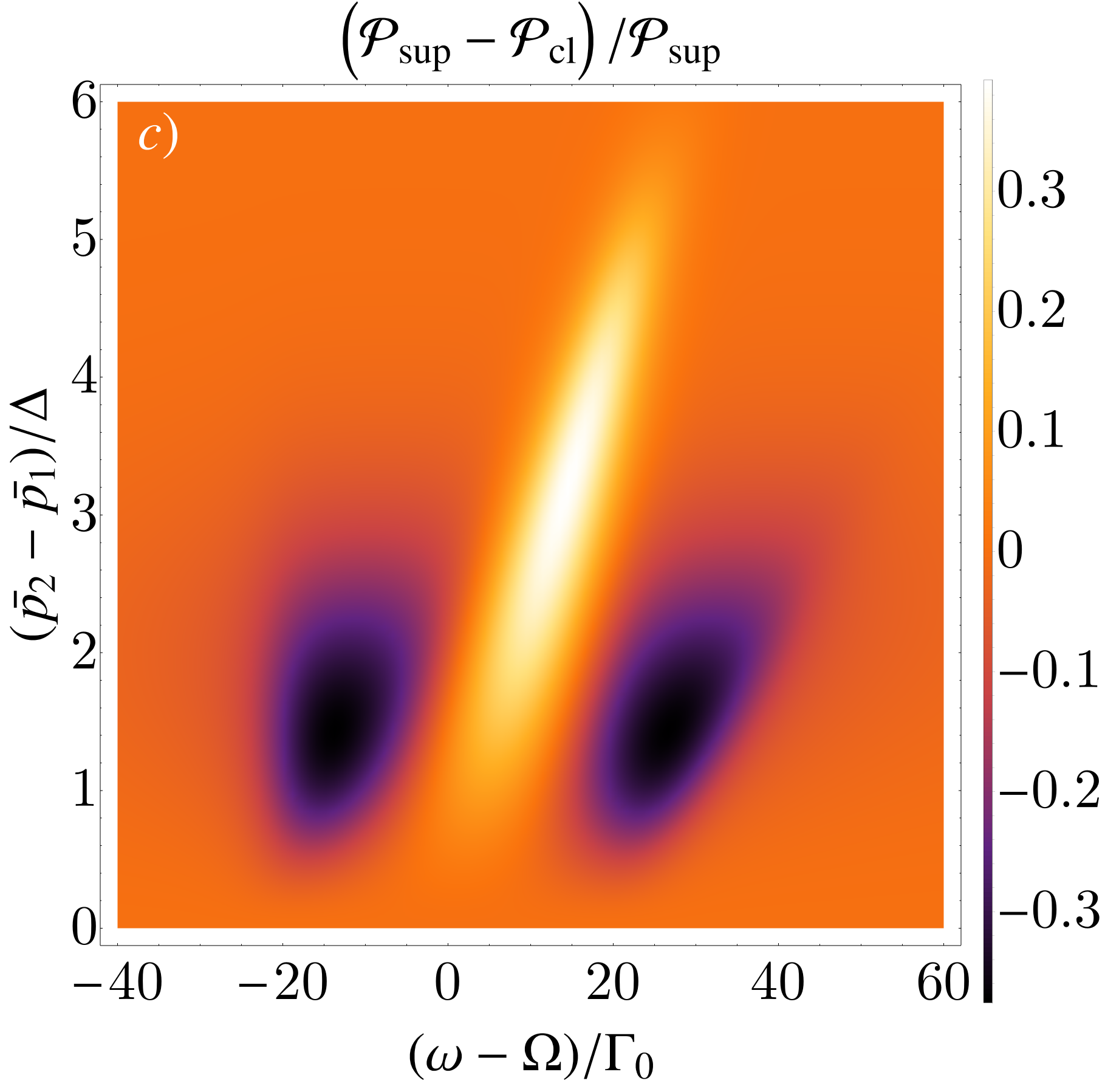}
	\caption{\label{plot3} 
 A comparison between shapes of the emission spectrum $ \mathcal{P}_{\parallel} (\omega)$ associated with a coherent superposition ($\mathcal{P}_{\text{sup}}$) and an incoherent classical mixture	($\mathcal{P}_{\text{cl}}$) of momentum wave packets a) The transition line for an atom initially prepared in a superposition of two momentum wave packets as a function of emitted photon's frequency and difference between wave packets' momenta. 
 The two-peak structure stemming from two distinct Doppler shifts is clearly visible.
 b) Absolute difference of emission probabilities between a superposition and a classical mixture of momentum wave packets as a function of the frequency of the emitted photon and difference between wave packets' momenta.
 The difference is most pronounced in regimes where wave packets overlap.
 c) Relative difference of emission probabilities between a superposition and a classical mixture of momentum wave packets as a function of the frequency of emitted photon and difference between the wave packets' momenta.
 It can be seen that the quantum contribution is largest in between the two transition peaks.
 This suggests that a postselection of final measured states of center-of-mass motion may increase the general visibility of the quantum Doppler effect.}
\end{figure*}
The denominator can be expanded up to first order in $\hbar \Omega / m c^2$ and up to the second order in $p^2 / m^2 c^2$ yielding
\begin{align}
\frac{1}{|\lambda'(\omega_0)|} \approx 1 + \frac{p}{m c} \cos \Theta + \frac{p^2}{m^2 c^2} \cos^2 \Theta - \frac{\hbar \Omega}{ m c^2},
\end{align}
and finally
\begin{align}
\frac{\eta(\omega_0)}{|\lambda'(\omega_0)|} &\approx  \frac{8 \pi}{3}  \sin \Theta \Big[ \ \Xi_0 (\Theta,\Phi) \left( 1- \frac{3}{2} \frac{\hbar \Omega}{2 m c^2} \right) \nonumber \\
& \quad  + \Xi_1 (\Theta,\Phi)\frac{p}{ m c} + \Xi_2 (\Theta,\Phi) \frac{p^2}{ 2 m^2 c^2} \Big],
\end{align}
where
\begin{align} 
\Xi_0 (\Theta,\Phi) &\equiv \frac{3}{8 \pi}  \left( 1 - \sin^2 \Theta \cos^2 \Phi\right),  \\
\Xi_1 (\Theta,\Phi)  &\equiv \frac{3}{4 \pi} \cos \Theta \left( 1 - 2 \sin^2 \Theta \cos^2 \Phi \right),  \\
\Xi_2 (\Theta,\Phi) &\equiv  \frac{3}{16 \pi}  
 \left[ 6 \cos 2 \Theta + 5 \cos^2 \Phi \left( \cos 4 \Theta - \cos 2 \Theta \right)\right]\!.
\end{align}
Substituting this expression into Eq.~\eqref{angular} yields the angular distribution
\begin{align} \label{fulang2}
\frac{\Gamma(\Theta,\Phi)}{\Gamma_0} &=   \Xi_0 (\Theta,\Phi) \left( 1 - \frac{3}{2}\frac{ \hbar \Omega}{  m c^2} \right)   \nonumber \\
&\quad +  \frac{1}{m c} \Xi_1 (\Theta,\Phi) \int \dd p \ p |\psi(p)|^2  \nonumber \\
&\quad + \frac{1}{2 m^2 c^2} \Xi_2 (\Theta,\Phi)  \int \dd p \ p^2 |\psi(p)|^2.
\end{align}
The difference between coherent and incoherent cases is then given by 
\begin{align}\label{angdiff2}
\frac{\Gamma_{\text{sup}} (\Theta,\Phi) \!-\Gamma_{\text{cl}} (\Theta,\Phi) }{\Gamma_0} = \Xi_1 (\Theta,\Phi) \ \delta_{\rm Q} + \Xi_2 (\Theta,\Phi) \ \gamma_{\rm Q}^{-1}. 
\end{align}

Integrating $\Gamma(\Theta,\Phi)$ over $\Theta$ and $\Phi$ yields the total transition rate
\begin{align}\label{tot}
\Gamma = \Gamma_0 \left(1 - \frac{3 \hbar \Omega}{ 2 m c^2} - \frac{1}{2 m^2 c^2} \int \dd p \ p^2 |\psi(p)|^2  \right).
\end{align}
If $\psi(p)$ is a wave packet well localized at $p_0$, then
\begin{align}\label{tot2}
\Gamma = \Gamma_0 \left(1 - \frac{3 \hbar \Omega}{ 2 m c^2} - \frac{p_0^2}{2 m^2 c^2} \right).
\end{align}
One can also immediately see that the quantum time dilation manifests in the total transition rate:
\begin{align}\label{corrections4}
\frac{\Gamma_{\text{sup}} -\Gamma_{\text{cl}}}{\Gamma_0} &= \gamma_{\rm Q}^{-1}.
\end{align}
The first term in Eq.~\eqref{fulang2} corresponds to the distribution of dipole radiation for an atom at rest which when integrated over $\Theta$ and $\Phi$ gives the transition rate $\Gamma_0$.
The second term is a correction linear in $p$ that associated with a Doppler shift. 
This term vanishes when integrated over $\Theta$ and $\Phi$, ensuring consistency with special relativity as the total transition rate is $\Gamma = \Gamma_0 / \gamma (p)$, which when expanded in $p$ is seen to have zero contribution linear in momentum.

However, terms linear in momentum modify the angular distribution of radiation, manifesting as a pattern distinctively different than that of the dipole radiation distribution.
The magnitude of this quantum correction depends on $K_1$ (i.e.\ $\delta_{\rm Q}$), which is an explicit function of the parameters characterizing the atomic wave functions and surprisingly vanishes if the atom moves in an equally weighted superposition.

Also of interest is how momentum coherence affects the shape of the atomic emission line, which can be probed through spectroscopic methods.
For a plane wave characterized by a wave vector $\bm{k}$, the probability of emission is given by
\begin{align}
 \mathcal{P}\left( \bm{k}\right) &= \lim_{t \rightarrow \infty} \sum_{\xi} \int \dd \bm{p} \ \left| \beta_{\bm{k}, \xi} \left( \bm{p},t \right) \right|^2,
\end{align}
which utilizing Eq.~\eqref{defeqs} can be cast into form
\begin{align}
 \mathcal{P}\left( \bm{k}\right) &= \frac{3 \Gamma_0}{16 \pi^2} \int \dd \bm{p}  \left| \psi \left( \bm{p} \right) \right|^2 \sum_{\xi} \frac{g_{\bm{k}, \xi}^2 (\bm{p}) / d^2}{\lambda^2 (\omega) + \Gamma^2 (\bm{p}) /4}.
\end{align}
By expanding $\lambda (\omega)$, $g_{\bm{k}, \xi}^2 (\bm{p}) / d^2$ and $\Gamma^2 (\bm{p})$ up to second order in momentum, under the assumption that the emission line is measured perpendicular to the direction of motion, one finds
\begin{align}\label{tranalsup01}
\mathcal{P}_{\perp} (\omega)  &=\frac{3}{8 \pi}  
\int \dd p \, \left| \psi \left( p \right) \right|^2  \nonumber \\
&\quad\times\frac{ \left( 1 -\frac{3}{2}\frac{ p^2}{m^2 c^2 } \right) \Gamma_0 / 2 \pi  }{\left[ \omega - \Omega \left(1 - \frac{1}{2}\frac{ p^2}{m^2c^2}\right) \right]^2 + \frac{\Gamma_0^2}{4} \left( 1-  \frac{ p^2}{m^2c^2} \right) }.
\end{align}
On the other hand, in the case of photons measured parallel to the direction of motion, one obtains
\begin{align}\label{tranalsup0}
\mathcal{P}_{\parallel} (\omega) &=\frac{3}{8 \pi}  
\int \dd p \, \left| \psi \left( p \right) \right|^2  \nonumber \\
&\quad\times\frac{ \left( 1 + 3 \frac{ p}{m c } \right) \Gamma_0 / 2 \pi  }{\left[ \omega - \Omega \left(1+ \frac{ p}{mc}\right) \right]^2 + \frac{\Gamma_0^2}{4} \left( 1+ 2 \frac{ p}{mc} \right) }.
\end{align}
 In Fig.~\ref{plot3} we compare the parallel emission spectrum $\mathcal{P}_{\parallel} (\omega)$  for coherent superposition and incoherent classical mixtures of momentum wave packets.
           
\bibliography{QuantumTimeDilation}

\end{document}